\documentclass[journal]{IEEEtran}
\usepackage{graphicx}
\usepackage{amsmath}
\usepackage{algorithm}
\usepackage{hyperref}
\usepackage{booktabs} 
\usepackage{tabularx} 

\usepackage{booktabs} 
\usepackage{tabularx} 
\usepackage{caption}  
\usepackage{mathtools}
\usepackage{amssymb}
\usepackage{amsmath}
\usepackage{multirow}
\usepackage{tabularx} 
\usepackage{array}
\usepackage{makecell}
\usepackage{textgreek}
\usepackage{comment}
\usepackage{bm}
\usepackage{booktabs}
\usepackage{color}
\usepackage{tabularx}
\usepackage{algorithmicx}
\usepackage{algpseudocode}
\usepackage{subcaption}
\usepackage[normalem]{ulem}
\usepackage{xcolor}
\usepackage{colortbl} 
\usepackage{multirow} 
\usepackage{makecell} 

\algnewcommand\Input{\item[\textbf{Input:}]}  
\algnewcommand\Output{\item[\textbf{Output:}]}  
\usepackage{graphicx} 
\begin{document}

\title{Can Movable Antenna-enabled Micro-Mobility Replace UAV-enabled Macro-Mobility? A Physical Layer Security Perspective}
 
\author{Kaixuan Li, Kan Yu,~\IEEEmembership{Member,~IEEE}, Dingyou Ma,~\IEEEmembership{Member,~IEEE}, Yujia Zhao, Xiaowu Liu, Qixun Zhang,~\IEEEmembership{Member,~IEEE}, and Zhiyong Feng,~\IEEEmembership{Senior Member,~IEEE}
\thanks{This work is supported by the National Natural Science Foundation of China with Grants 62301076 and 62401077, Fundamental Research Funds for the Central Universities with Grant  24820232023YQTD01, National Natural Science Foundation of China with Grants 62341101 and 62321001, Beijing Municipal Natural Science Foundation with Grant L232003, and National Key Research and Development Program of China with Grant 2022YFB4300403.
}

\thanks{K. Li is with the School of Computer Science, Qufu Normal University, Rizhao, P.R. China. E-mail: lkx0311@126.com;}
\thanks{K. Yu (\emph{the corresponding author}) is with the Key Laboratory of Universal Wireless Communications, Ministry of Education, Beijing University of Posts and Telecommunications, Beijing, 100876, P.R. China. E-mail: kanyu1108@126.com;}
\thanks{D. Ma is with the Key Laboratory of Universal
	Wireless Communications, Ministry of Education, Beijing University of Posts and Telecommunications, Beijing, 100876, P.R. China. E-mail: dingyouma@bupt.edu.cn;}
\thanks{Y. Zhao is with the Key Laboratory of Universal
	Wireless Communications, Ministry of Education, Beijing University of Posts and Telecommunications, Beijing, 100876, P.R. China. E-mail: zhaoyj0318@bupt.edu.cn;}
\thanks{X. Liu is with the School of Computer Science, Qufu Normal University, Rizhao, P.R. China. E-mail: liuxw@qfnu.edu.cn;}
\thanks{Q. Zhang is with the Key Laboratory of Universal
Wireless Communications, Ministry of Education, Beijing University of Posts and Telecommunications, Beijing, 100876, P.R. China. E-mail: zhangqixun@bupt.edu.cn;}
\thanks{Z. Feng is with the Key Laboratory of Universal Wireless Communications, Ministry of Education, Beijing University of Posts and Telecommunications, Beijing, 100876, P.R. China. E-mail: fengzy@bupt.edu.cn.}


}

\markboth{IEEE Transactions on Mobile Computing,~Vol.~, No.~, 2025}%
{Shell \Baogui Huang{\textit{et al.}}: Shortest Link Scheduling Under SINR}
\maketitle
\begin{abstract}
This paper investigates the potential of movable antenna (MA)-enabled micro-mobility to replace UAV-enabled macro-mobility for enhancing physical layer security (PLS) in air-to-ground communications. While UAV trajectory optimization offers high flexibility and Line-of-Sight (LoS) advantages, it suffers from significant energy consumption, latency, and complex trajectory optimization. Conversely, MA technology provides fine-grained spatial reconfiguration (antenna positioning within a confined area) with ultra-low energy overhead and millisecond-scale response, enabling real-time channel manipulation and covert beam steering.
To systematically compare these paradigms, we establish a dual-scale mobility framework where a UAV-mounted uniform linear array (ULA) serves as a base station transmitting confidential information to a legitimate user (Bob) in the presence of an eavesdropper (Eve). We formulate non-convex average secrecy rate (ASR) maximization problems for both schemes: 1) MA-based micro-mobility: Jointly optimizing antenna positions and beamforming (BF) vectors under positioning constraints; 2) UAV-based macro-mobility: Jointly optimizing the UAV's trajectory and BF vectors under kinematic constraints. 
Extensive simulations reveal distinct operational regimes: MA micro-mobility demonstrates significant ASR advantages in low-transmit-power scenarios or under antenna constraints due to its energy-efficient spatial control. Conversely, UAV macro-mobility excels under resource-sufficient conditions (higher power, larger antenna arrays) by leveraging global mobility for optimal positioning. The findings highlight the complementary strengths of both approaches, suggesting hybrid micro-macro mobility as a promising direction for balancing security, energy efficiency, and deployment complexity in future wireless networks.
\end{abstract}

\IEEEpeerreviewmaketitle

\section{Introduction}\label{sec:introduction}
Unmanned Aerial Vehicles (UAVs) have emerged as pivotal platforms for next-generation wireless communication, offering unique advantages such as rapid deployment, flexible altitude control, and high-probability line-of-sight (LoS) links. These attributes make UAVs ideal for applications ranging from IoT data harvesting and emergency response to coverage extension \cite{Meng2024UAVSurvey,Xiao2022Survey}. However, the inherent broadcast nature of wireless channels renders UAV communications highly susceptible to eavesdropping attacks, posing significant threats to information confidentiality. Physical layer security (PLS), exploiting the intrinsic randomness of wireless channels to achieve secure transmission without relying on cryptographic keys, has thus become a critical technique for safeguarding UAV communications \cite{He2025Review,Wang2022Physical}. A fundamental objective in PLS design is the maximization of the secrecy capacity (or secrecy rate), defined as the difference between the achievable rates of the legitimate link and the eavesdropper (Eve)'s link.

To enhance secrecy performance, the predominant strategy in UAV-aided secure communication involves the joint optimization of the UAV's flight trajectory and beamforming (BF) \cite{Zhang2024Sensing}. By dynamically planning the UAV's path and adjusting BF matrix, this approach aims to favorably shape the legitimate channel while degrading the quality of eavesdropping channel. While demonstrably effective in boosting secrecy rates, this paradigm suffers from several inherent and significant limitations that hinder practical deployment:
\begin{itemize}
	\item \emph{Substantial energy consumption}: UAV propulsion dominates onboard energy use. Security driven trajectory maneuvers rapidly drain battery reserves, critically curtailing endurance. Moreover, secrecy rate maximization fundamentally conflicts with propulsion energy minimization.
	\item \emph{Latency and adaptability deficits}: Computing and executing new trajectories introduces significant latency, under which evolving channel conditions render optimized trajectories obsolete, undermining sustained security;
	\item \emph{Compromising flight pattern alterations}: Conspicuous trajectory adjustments intended to evade Eves may inadvertently reveal the UAV's position or intention, heightening vulnerability.
\end{itemize}

Recently, movable antenna (MA) technology has emerged as a revolutionary hardware solution offering unprecedented spatial degrees of freedom (DoF) for fine-grained channel manipulation  \cite{Zhu2024Movable,Zhu2025Tutorial}. An MA enables the physical position of a transmit/receive antenna element to be dynamically adjusted within a confined region on a platform (e.g., a linear rail or 2D surface on the UAV). By precisely controlling the antenna's location (typically on centimeter or millimeter scales), the position-dependent channel response (amplitude and phase) to both the legitimate receiver and potential Eves can be actively engineered, enabling real-time beam pattern and phase front shaping. Crucially, MA technology promises transformative advantages over trajectory manipulation:
\begin{itemize}
	\item \emph{Electromechanical efficiency}: Repositioning lightweight antenna elements incurs negligible energy overhead compared to UAV propulsion, fundamentally contrasting electromechanical actuation with platform movement \cite{Zhu2025Tutorial,Li2025Movable};
	\item \emph{Ultra-fast response}: Millisecond-scale antenna relocation enables real-time channel tracking--a critical advantage for dynamic environments where second-scale trajectory adjustments exceed channel coherence times \cite{Zhu2024Movable,zhu2023modeling};
	\item \emph{Covert beam steering}: MA-enabled beam reorientation avoids kinematic signatures associated with platform maneuvering \cite{Zhu2024Movable}.
	\end{itemize}
The integration of MAs on UAV platforms establishes a hierarchical control paradigm: UAV macro-mobility provides coverage, while MA micro-positioning enables real-time channel state manipulation with minimal energy overhead. This dual-scale approach fundamentally overcomes trajectory-based PLS limitations. Nevertheless, the capability of UAV-mounted MAs to supersede trajectory optimization for robust PLS remains experimentally unvalidated.
This work pioneers the exploration of UAV-mounted MAs for PLS, introducing a paradigm shift from conventional BF optimization. As a result, \emph{``Can joint optimization of antenna positioning and transmit BF matrix achieve comparable or superior secrecy performance to conventional joint trajectory-BF design, while drastically reducing computational overhead and propulsion energy consumption?''} needs to be answered.

To systematically quantify performance boundaries between spatial control strategies, we develop a dual-scale mobility framework for an air-to-ground scenario where a UAV-mounted uniform linear array (ULA) structure (FPA for UAV trajectory control or MA for antenna positioning) serves as an aerial base station and transmits the confidential information to a legitimate receiver (Bob), in the presence of an Eve. Both Bob and Eve equip with fixed antennas. 
This framework quantifies the performance gap between UAV-based macro-mobility and MA micro-positioning, enabling a systematic evaluation of their respective advantages and applicability in communication scenarios under varying system conditions.
To our knowledge, this constitutes the first systematic quantification of macro-micro mobility substitutability for. The main contributions of this paper can be summarized as follows.
\begin{itemize}
	\item We establish a novel air-to-ground communication model to systematically compare MA-enabled micro-mobility and UAV-enabled macro-mobility for PLS. Furthermore, non-convex average secrecy rate (ASR) maximization problems for both mobility paradigms are formulated, namely joint optimization of antenna positioning and BF vectors under spatial constraints for MA micro-mobility, while joint optimization of UAV trajectory and BF under kinematic constraints for UAV macro-mobility;
	\item We develop an efficient AO framework to decompose complex joint optimizations into tractable subproblems: 1) Integrated projected gradient ascent (PGA) with simulated annealing (SA) to optimize antenna positioning and BF for MA micro-mobility, avoiding local optima; 2) Combined successive convex approximation (SCA) with interior-point methods to handle trajectory-BF coupling For UAV macro-mobility. Comprehensive computational complexities are analyzed;
	\item Simulated results demonstrate that MA micro-mobility outperforms in low-power scenarios (less than 0.1W) and antenna-constrained settings (optimal at 4 and 5 antennas), leveraging energy-efficient spatial reconfiguration; 2) UAV macro-mobility excels under high-power regimes (greater than 1W) and with large antenna arrays, utilizing global mobility for optimal positioning. These findings highlight the adaptive suitability of each scheme under distinct application scenarios, offering critical insights into their respective advantages and optimal deployment conditions.
\end{itemize}

The remainder of the paper is organized as follows: The related works on UAV and MA assisted technologies are demonstrated in Section \ref{sec:related work}. In Section \ref{sec:network model}, the system model is introduced, and the ASR maximization problems and the performance gap function are formulated. In Section \ref{sec:opt}, communication duration of confidential information is optimized. Simulations are conducted to demonstrate how the performance gap between the two schemes evolves under different system parameters. Finally, conclusions and future works are discussed in Section \ref{sec:conclusion}.

Notations: In this paper, $ (\cdot)^\mathbf{H}$ and  $ (\cdot)^\mathbf{T}$ denote the Hermitian (conjugate transpose) and transpose operations, respectively. $\textbf{tr}(\mathbf{X})$ represents the trace with diagonal elements $\mathbf{X}$. $\nabla$ denote the gradient operator.

\section{Related Works}\label{sec:related work}
PLS methods are gaining significant attention in wireless networks due to their potential for safeguarding sensitive information without relying on cryptographic techniques. Among emerging PLS research trends, UAV trajectory design, MA positioning, and secure BF design have attracted considerable interest for their ability to exploit spatial degrees of freedom to enhance secrecy performance. In this section, we comprehensively review representative PLS schemes leveraging both UAV macro-trajectory optimization and MA micro-positioning to improve the whole system's performance.

\begin{table*}[t]
\caption{\small Main researches on MA/UAV-enabled PLS design}
\label{tab:mobility_studies}
\centering
\arrayrulecolor{black}
\arrayrulewidth=0.5pt
\renewcommand{\arraystretch}{1.5}
\begin{tabular}{|
>{\columncolor[HTML]{DEEAF4}\centering\arraybackslash}m{2cm}|
>{\columncolor[HTML]{DEEAF4}\centering\arraybackslash}m{2.5cm}|
>{\columncolor[HTML]{ECEDEE}\centering\arraybackslash}m{4cm}|
>{\columncolor[HTML]{DEEAF4}\centering\arraybackslash}m{2.5cm}|
>{\columncolor[HTML]{DEEAF4}\arraybackslash}m{4.8cm}|}
\hline
\textbf{\cellcolor[HTML]{9DC2E3}Mobility Type} & \textbf{\cellcolor[HTML]{9DC2E3}Reference} & \textbf{\cellcolor[HTML]{9DC2E3}Techniques} & \textbf{\cellcolor[HTML]{9DC2E3}Metrics} & \textbf{\cellcolor[HTML]{9DC2E3}Key Contributions} \\ \hline

& \emph{Zhang et al.} \cite{Zhang2023Achieving} 
& 3D trajectory design of friendly jamming UAV 
& Secrecy region 
& Investigated the effect of UAV’s 3D position and jamming power on expanding the secrecy region. \\ \cline{2-5}

& \emph{Zhang et al.} \cite{Zhang2025Robust} 
& 3D trajectory design of UAV's trajectory, transmit power, jamming BF and user scheduling 
& Secrecy rate 
& Jamming signals were utilized to degrade the eavesdropper's channel, and their BF is jointly designed for optimal interference. \\ \cline{2-5}
 
\multirow{2}{*}{\centering \textbf{macro-mobility}}
& \emph{Zhang et al.} \cite{Zhang2024Sensing} 
& Joint design of UAV's 2D trajectory, transmit power and RIS phase shift co-design 
& ASR 
& Proposed a joint UAV-RIS framework with alternating optimization to achieve near-optimal secrecy rate under perfect CSI. \\ \cline{2-5}

& \emph{Xiu et al.} \cite{Xiu2025Improving} 
& Joint design of UAV's 2D trajectory and ISAC BF 
& ASR 
& Proposed a novel autonomous aerial vehicle secure communication system with ISAC.\\ \cline{2-5}

& \emph{Tarekegn et al.} \cite{Tarekegn2025} 
& Multi-UAV 3D trajectory optimization  
& Communication coverage, network throughput 
& Developed a federated deep reinforcement learning scheme to balance communication coverage and throughput. \\ \cline{2-5} \hline

& \emph{Ma et al.} \cite{Ma2024MIMO} 
& 6D MA arrays 
& Channel capacity 
& Demonstrated significant capacity gains over FPA-MIMO, especially in rich multipath environments. \\ \cline{2-5}

& \emph{Shao et al.} \cite{Shao2025Dis} 
& Joint design of directional sparse matrix and  6D MA placement design
& Detection error rate, normalized mean squared error 
& First identified and exploited directional sparsity in MA-assisted systems, improving detection and estimation accuracy. \\ \cline{2-5}

\multirow{2}{*}{\centering \textbf{ Micro-mobility}} 
& \emph{Tang et al.} \cite{Tang2025Secure} 
& Joint design of transmit precoding, AN covariance and 2D MA positioning 
& Secrecy rate 
& Pioneered the evaluation of PLS performance in MA-assisted systems under optimized spatial configurations. 
\\ \cline{2-5}

& \emph{Hu et al.} \cite{Hu2024Secure} 
& Joint design of the BF and MA's position of transmitter 
& Secrecy rate 
& Investigated secure communication from an MA-equipped transmitter to a single-antenna receiver under multiple Eves.
\\ \cline{2-5}

& \emph{Feng et al.} \cite{Yujia2024} 
& Joint design of the BF and MA's position of transmitter 
& Secrecy rate 
& Moving partial antennas may yield better results than moving all antennas.
\\ \cline{2-5}

& \emph{Wang et al.} \cite{Wenxu}
& Joint design of the BF of MA and FPA, and MA's position of transmitter  
& Secrecy rate 
& MA functions as an interference generator to suppress the eavesdropping channel.
\\ \hline

\centering \textbf{Macro- \& Micro-mobility} 
& \emph{Liu et al.} \cite{Liu2025UAVMA} 
& Joint design of UAV trajectory, MA positions, transmit BF 
& Throughput 
& Demonstrated the feasibility of hybrid mobility strategies in boosting secure transmission capacity. \\
 \hline

\end{tabular}
\vspace{0.5cm}
\raggedright
\end{table*}

\subsection{UAV macro-mobility based PLS design}
UAV-enabled macroscopic mobility has emerged as a compelling solution for PLS enhancement, owing to its ability to flexibly control trajectories and establish favorable LoS links.
 
By planning and tuning the two-dimensional (2D) trajectory of the UAV with integrated and communication systems, it was shown in \cite{Chai2024Sys} that UAV can significantly improve the network capacity. 
Extending into three-dimensional (3D) space, the trajectory and transmission power of the jammer UAV were carefully designed to maximize the security region \cite{Zhang2023Achieving}.
Similar works were done \cite{Wang2020Physical,Yu2025SuRLLC,Wang2021Robust}. 
Unlike BF design schemes for confidential signals, \emph{Zhang et al.} exploited base station-generated jamming signals to degrade the eavesdropping channel and optimized their waveform design accordingly \cite{Zhang2025Robust}.
In addition, the sensing signal was exploited by \emph{Zhang et al.} for user localization and channel state information (CSI) estimation, leading to the formulation of a joint design problem for UAV trajectory and receive BF \cite{Zhang2024Sensing}. Focusing on the ISAC secure communication system, where the UAV is equipped with a ULA, the average communication secrecy rate was maximized by optimizing the UAV trajectory and ISAC-based BF vectors, as proposed in \cite{Xiu2025Improving}. 
Diverging from the aforementioned algorithm, in \cite{Tarekegn2025}, \emph{Tarekegn et al.} proposed a novel multi-UAV trajectory control and fair communication framework, by leveraging federated multi-agent deep reinforcement learning (DLA) to autonomously optimize 3D flight paths, which concurrently maximized ground user data rates and network coverage. In both ULA and UPA configurations, \emph{Sun et al.} derived closed-form lower-bound equations for the ASR after conducting a thorough investigation of the three-dimensional antenna gain characteristics \cite{Sun2021Secure}. 

To sum up, these studies reveal the superiority of UAV macro-scale mobility in achieving substantial performance enhancements across secure communication, network coverage, and system throughput optimization.

\subsection{MA micro-mobility based PLS design}
In recent years, numerous studies have demonstrated the significant advantages of MA over fixed-position antennas, specifically with regard to improved spatial multiplexing capabilities, interference suppression, flexible BF design, and signal power enhancement \cite{Zhu2024Movable}. 
 
In \cite{Ma2024MIMO}, an MA-MIMO system was proposed to maximize channel capacity through the joint optimization of the transmit and receive MA positions and the transmit signal covariance matrix, demonstrating significant advantages over conventional FPA-MIMO systems, particularly in multipath-rich environments.
In comparison, six-dimensional MAs (6DMA) offer significantly enhanced spatial degrees of freedom, which can
be separately adjusted in terms of both 3D positions and 3D rotations, subject to practical movement constraints. In \cite{Shao2025Dis}, \emph{Shao et al.} identified the directional sparsity property of 6DMA systems, under which a practical three-stage protocol for CSI estimation was established, and the user's instantaneous channel was optimized. 

However, the threat posed by the existence of eavesdropping parties to the system was ignored by the above-mentioned research. To alleviate the risk of confidential signal interception, \emph{Tang et al.} employed the assumption of perfect CSI to simplify the design of the transmit precoding matrix, artificial noise covariance, and MA positions in pursuit of maximizing the secrecy rate \cite{Tang2025Secure}. Building upon this idea, in \cite{Liu2025UAVMA}, \emph{Liu et al.} explored a UAV-mounted linear MA system under the pretense of perfect CSI, where the goal was to enhance throughput capacity by jointly optimizing the transmit BF, the UAV trajectory, and the positions of the MA array elements. 
Considering the secure transmission with the CSI of eavesdropping parties being unknown, in \cite{Hu2024Movable}, \emph{Hu et al.} derived an approximate expression of the secrecy outage probability by using the Laguerre series to approximate the close distribution of the received power gain of eavesdropping parties. Moreover, a joint optimization of the BF at the transmitting end and the antenna position was done for minimizing the secrecy outage probability. 
Furthermore, in \cite{Hu2024Secure}, they jointly designed the transmit BF and MA's positions within the budget constraints, aiming to enhance the secrecy rate. A similar work was presented in \cite{Yujia2024}, \emph{Feng et al.} demonstrated that moving partial antennas may yield better results than moving all antennas for enhancing the secrecy rate. In \cite{Wenxu}, \emph{Wang et al.} adopted MA functions as an interference generator, replacing the traditional FPA-based interference generation scheme. To overcome the analytical challenges posed by unknown Eves' locations, an equivalent model was proposed, which approximated the impact of multiple static Eves using a single virtual Eve equipped with an MA array, thereby capturing the spatial reconfigurability of MA \cite{Yujia2}.

\begin{table}[!htb]
	\caption{\small Key symbols and meanings in this paper}
	\centering
	\renewcommand{\arraystretch}{1.2} 
	\label{tab:network_symbols}
	\begin{tabular}{lll}
		\toprule
		Symbols & Meanings & Pages\\
		\midrule
		$H$ & UAV flight altitude & 4\\
		$M$ & number of antennas & 4\\
		$T$ & total mission duration  & 4  \\
		$N$ & number of time slots   & 4 \\ 
		$\mathbf{q}_u$ & UAV position & 4\\
		$\mathbf{p}^{\mathrm{Loc}}_m$ & $m$-th  antenna position    & 4 \\
		$d^{\rm FPA}_{\text{fix}}$ & FPA's antenna spacing   & 4 \\
		$d_{\min}^{\mathrm{MA}}$ & minimum inter-MA spacing & 4 \\
		$L_{\max}^{\rm MA}$ & maximum MA-spacing between two time slots & 5\\
		$\theta_i$ & elevation angle from UAV to user $i$    & 4\\
		$\phi_i$ & azimuth angle from UAV to ground user $i$   & 4 \\ 
		$d_{u,i}$ & distance between UAV and ground user $i$   & 4 \\
		$\mathbf{a}_i$ & steering vectors for ground user $i$  & 4\\
		$\mathbf{h}_i$ & channel vector from UAV to ground user $i$  & 4 \\
		$\mathbf{w}$ & transmitting BF vector    & 4\\
		$\sigma_i^2$ & noise power of user $i$    & 5\\
		$x^{m,\rm Loc}_{\min}, x^{m,\rm Loc}_{\max}$ & $m$-th antenna positioning bounds  & 5  \\
		$P_{\max}$ &  maximum communication power  & 5  \\
		$\gamma_i$ &  SINR of ground user $i$  & 5  \\
		$\tau$ & the secrecy rate $i$  & 5  \\
		$v_u,a_u$ & velocity and acceleration of UAV   & 6  \\
		$v_{\max},a_{\max}$ & maximum velocity and acceleration of UAV   & 6  \\
		\bottomrule
	\end{tabular}
\end{table}

Table~\ref{tab:mobility_studies} provides a comprehensive overview of prominent PLS designs leveraging UAV macro-mobility and MA micro-mobility. Nevertheless, prior work lacks a systematic comparative analysis quantifying performance difference between MA micro-mobility and UAV macro-mobility paradigms.

\section{Network Model and Preliminaries}\label{sec:network model}
\begin{figure}[!ht]
    \centering
    \includegraphics[width=9cm]{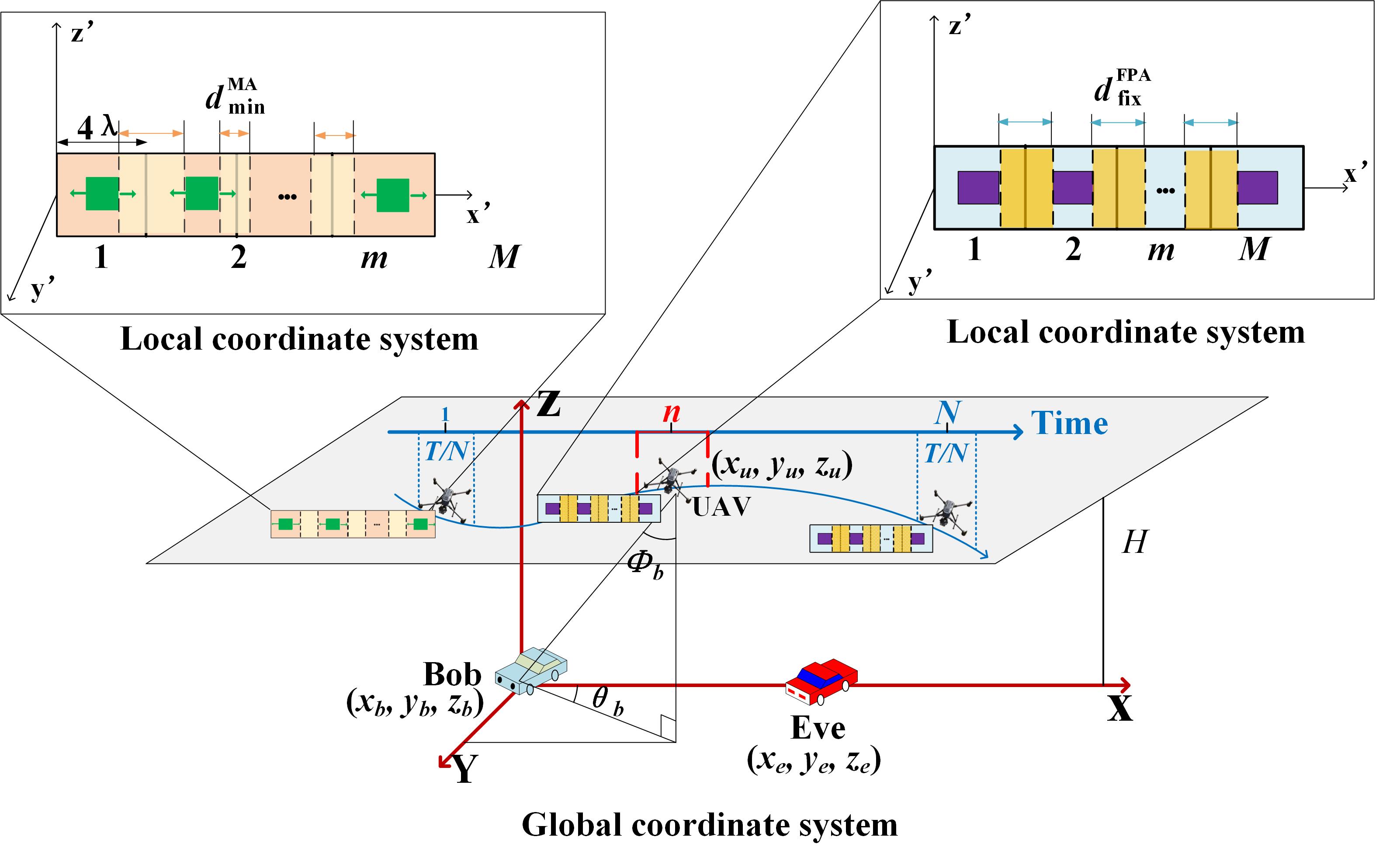}
\caption{\small System model}
\label{fig:system_model}
\end{figure}

As illustrated in Fig. \ref{fig:system_model}, 
an air-to-ground communication framework is formulated to investigate micro-macro mobility impacts on PLS. Specifically, a UAV equipped with a $M$ transmitting antennas featured by ULA serving as aerial base station sends confidential information to the Bob, in the presence of an passive Eve. Both the Bob and the Eve employ single receiving antennas with fixed positions at $(0,0,0)$ and $(x_e,y_e,0)$, respectively.
To more precisely capture the temporal system evolution, the total mission duration $T$ is partitioned into $N$ time slots of duration $\Delta_t = T/N$, indexed by $n$.  For UAV macro-mobility characterization, UAV altitude is fixed at $H$m with horizontal trajectory, and $M$ transmitting antennas are regarded as FPAs. As a result, the position of UAV at $n$-th time slot is $\mathbf{q}_u[n] = (x_u[n], y_u[n], H)$. To mitigate mutual coupling between antenna elements, the antennas are typically separated by a distance of $d^{\rm FPA}_{\text{fix}}$ ($\geq1/2\lambda$). For MA micro-mobility characterization, UAV maintains hovering at altitude $H$m, and each of $M$ transmitting antennas adjusts position within $[0, 4\lambda]$, where $\lambda$ denotes the wavelength. In this case, the $m$-th antenna position can be represented as $\mathbf{p}^{\mathrm{Loc}}_m[n] = (x^{\mathrm{Loc}}_m[n], y^{\mathrm{Loc}}_m[n], 0)$ for $m \in \{1,\ldots,M\}$. The minimum inter-antenna spacing among movable elements is denoted by $d_{\min}^{\mathrm{MA}}$. In addition, elevation $\theta_i[n]$ and azimuth $\phi_i[n]$ from UAV to node $i \in \{\mathrm{Bob}, \mathrm{Eve}\}$ are defined, and $\cos(\alpha_i[n]) =\sin (\theta_i[n])\cdot \cos (\phi_i[n])$ \cite{Sun2021Secure,Yuan2019Joint}.
The key symbols and variables used in this paper are summarized in Table \ref{tab:network_symbols} for clarity and ease of reference.

\subsection{Communication model}
To facilitate tractable analysis, the UAV-to-ground links are modeled as deterministic LoS channels, thereby explicitly excluding complexities arising from aerial scattering and blockage phenomena. Consistent with common practice in related literature \cite{Yan2025Securing,Chu2023Joint}, perfect knowledge of user positions and CSI is assumed at both Bob and Eve.
Let $\mathbf{w}[n] \in \mathbb{C}^{M \times 1}$ be the BF vector generated by UAV with $M$ transmitting antennas. Based on the conclusion of \cite{Ma2024MIMO,Chen2023Joint}, the received signal at Bob from UAV can be expressed as 
\begin{equation}
y_b[n] = \mathbf{h}_b[n]^{\rm \mathbf{H}} \mathbf{w}[n] s[n] + \sigma_b^2[n],
\end{equation}
where $\mathbf{h}_b[n] \in \mathbb{C}^{M \times 1}$ denotes the channel response from the UAV to the Bob at the $n$-th time slot, which is dependent on their instantaneous positions, $s[n]$ represents the transmitted signal with zero mean and unit power, and $\sigma_b^2[n]$ denotes the additive white Gaussian noise (AWGN) at the Bob.  
Similarly,the signal received at the Eve is given by
\begin{equation}
y_e[n] = \mathbf{h}_e[n]^{\rm \mathbf{H}} \mathbf{w}[n] s[n] + \sigma_e^2[n],
\end{equation}
where $\mathbf{h}_e[n] \in \mathbb{C}^{M \times 1}$ signifies the channel response from the UAV to the Eve at the $ n $-th time slot, and $\sigma_b^2[n]$ denotes the AWGN at the Eve.

For MA micro-mobility characterization, $\cos(\alpha_i) = \sin (\theta_i[0]) \cdot \cos (\phi_i[0]) $ holds, since the UAV hovers at its initial location. Accordingly, at the $n$-th time slot, the steering vectors from UAV to Bob and Eve are given by
\begin{equation}
\mathbf{a}_b[n] = 
\begin{bmatrix}
e^{j\frac{2\pi}{\lambda} x^{\rm Loc}_1[n] \cos\alpha_{b}[n]} ,&
\cdots &,
e^{j\frac{2\pi}{\lambda} x^{\rm Loc}_M[n] \cos\alpha_{b}[n]}
\end{bmatrix}^\mathbf{T},
\end{equation}
and
\begin{equation} 
\mathbf{a}_e[n] = 
\begin{bmatrix}
e^{j\frac{2\pi}{\lambda} x^{\rm Loc}_1[n] \cos\alpha_{e}[n]} &
\cdots &
e^{j\frac{2\pi}{\lambda} x^{\rm Loc}_M[n] \cos \alpha_{e} [n]}
\end{bmatrix}^\mathbf{T}, 
\end{equation}
respectively.

Given the elevated deployment of UAV and minimal ground obstructions, the air-to-ground channel exhibits a dominant LoS component\footnote{This justifies the adoption of a Rician fading model with high $K$-factor, where the deterministic LoS path governs channel gain while multipath effects induce minor fluctuations. Path loss follows the altitude-dependent logarithmic model in 3GPP TR 38.811, accounting for free-space propagation and environment-specific attenuation \cite{3gpp2018study,TR38}.}. Consequently, based on the conclusion of \cite{Wang2021Robust}, the path gain at the $n$-th time slot can be modeled using free-space path loss (FSPL) channel, which accurately captures LoS propagation conditions. This is,
\begin{equation}
g_i[n] = \frac{\beta_0}{d_{u, i}^{\alpha}[n]}, \quad i \in \{\mathrm{Bob}, \mathrm{Eve}\},
\end{equation}
where $\beta_0$ is the reference path loss at a standardized distance of 1 meter, and  $d_{u, i}[n]$ is the distance between the UAV to ground user $i$ at the $n$-th time slot.
Due to the dominated LoS propagation channel, the setting of  $\alpha = 2$ is reasonable. Then, we get
\begin{equation}
\mathbf{h}_{i}[n] = \sqrt{g_0} \, \mathbf{a}_i[n], \quad i \in \{\mathrm{Bob}, \mathrm{Eve}\}.
\end{equation}

Given the $n$-th time slot, the instantaneous signal-to-noise ratio (SNR) at the Bob and the Eve can be expressed as
\begin{equation} \label{eq:SNR_b}
\mathrm{\gamma }_b[n] = \frac{\left| \mathbf{h}_b^{\rm \mathbf{H}}[n] \mathbf{w}[n] \right|^2}{\sigma_b^2}
= \frac{ \beta_0 \left| \mathbf{a}_b^{\rm \mathbf{H}}[n] \mathbf{w}[n] \right|^2 }{ d_{u, b}^2[n] \, \sigma_b^2 },
\end{equation}
and
\begin{equation}\label{eq:SNR_e}
\mathrm{\gamma}_e[n] = \frac{\left| \mathbf{h}_e^{\rm \mathbf{H}}[n] \mathbf{w}[n] \right|^2}{\sigma_b^2}
= \frac{ \beta_0 \left| \mathbf{a}_e^{\rm \mathbf{H}}[n] \mathbf{w}[n] \right|^2 }{ d_{u, e}^2[n] \, \sigma_e^2 },
\end{equation}
respectively.

\emph{Average secrecy rate}: According to the Shannon theorem \cite{Mert2021}, the achievable rate (bits/s/Hz) between the UAV and the Bob at the $n$-th time slot is  $R_b[n] = \log(1 + \gamma_b[n])$. Similarly,
the achievable rate (bits/s/Hz) between the Alice and the Eve for decoding the confidential information at the $n$-th time slot is given by $R_e[n] = \log(1 + \gamma_e[n])$. Hence, the secrecy rate between the UAV and the Bob at the $n$-th time slot can be represented as
\begin{equation}
	\tau[n] = \left[R_b[n] - R_e[n] \right]^+, \\
\end{equation}
where $[x]^+ \triangleq \max(x, 0) $, and the ASR of the system over all time slots is given by
\begin{displaymath}
	 \frac{1}{N}\sum_{n=1}^{N} \tau[n],
\end{displaymath}
which represents a long-term secrecy performance of the systems across all time slots and provides a more comprehensive perspective for evaluating the overall secrecy performance rather than focusing on instantaneous state.

\subsection{Optimization problem formulation}
Motivated by micro-macro mobility impacts on PLS, different from the conventional
PLS design, this paper aims to answer the question of \emph{``Can joint optimization of antenna positioning and transmit BF matrix achieve comparable or superior secrecy performance to conventional joint trajectory-BF design?''}, by jointly optimizing the time-varying positions of the MA/UAV and the BF vectors of $M$ transmitting antennas.
Specifically, for the case of MA micro mobility, the variables are associated with the antenna positioning and BF vectors, and the optimization problem can be formulated as
\begin{subequations} \label{eq:P1}
\begin{align}
\text{P1:} \quad
&\max_{\mathbf{x}_m, \mathbf{w}} \quad  \frac{1}{N}\sum_{n=1}^{N} \tau[n] \\
\mbox{s.t.}\quad 
& \left| x^{\rm Loc}_m[n] - x^{\rm Loc}_m[n-1] \right| \leq L_{\max}^{\rm MA} , \forall m \in \{1, \dots, M\}, \notag\\ 
&\forall n \in \{1, \dots, N\}\label{eq:P1-2s} \\
& x^{\rm Loc}_m \in {\left[ x^{m,\rm Loc}_{\min},x^{m,\rm Loc}_{\max} \right]} ,  \label{eq:P1-3}\\
&\mathbf{ tr} \left( \mathbf{w} \mathbf{w}^{\rm H} \right) \leq P_{\max} \label{eq:P1-4}
\end{align}
\end{subequations}
where $\mathbf{x}_{\text{\it m}} = [x_1^{\rm Loc}, \ldots, x_M^{\rm Loc}]$ represents the x-axis of MA in local coordinate system; \eqref{eq:P1-2s} limits the displacement of the $m$-th antenna element within consecutive time slots to the maximum allowable movement distance $L_{\max}^{\rm MA}$; Constraint \eqref{eq:P1-3} restricts the antenna positioning to the designated feasible $\Psi_m$, while \eqref{eq:P1-4} enforces the transmit power budget with $P_{\max}$ denoting the upper bound.
For the case of UAV macro-mobility, all antenna elements become fixed. Then, the optimization problem can be written as
\begin{subequations} \label{eq:P2}
\begin{align}
\text{P2:} \quad
&\max_{\mathbf{q}_u, \mathbf{w}} \quad  \frac{1}{N}\sum_{n=1}^{N} \tau[n] \\
\mbox{s.t.}\quad 
& \left| v_u[n] \right| \leq v_{\max}, \quad \label{eq:P2-2} \\
& \left| a_u[n] \right| \leq a_{\max}, \quad \label{eq:P2-3} \\
& \mathbf{tr} \left( \mathbf{w} \mathbf{w}^{\rm H} \right) \leq P_{\max} \label{eq:P2-4}
\end{align}
\end{subequations}
Constraints \eqref{eq:P2-2} and \eqref{eq:P2-3} limit the UAV's velocity and acceleration vectors, respectively, and $v_{\max}$ and $a_{\max}$ denote their maximum allowable magnitudes.

However, Problem \eqref{eq:P1} and Problem \eqref{eq:P2} are non-convex optimization problems,
which are difficult to be solved directly, due to the following three reasons: 1) the operator of $[\cdot]^+$ is non-smoothness for the objective function; 2) the variables $\mathbf{x}_m$ (or $\mathbf{q}_u$) and ${\rm \mathbf{w}}$ are intricately coupling; 3) and they further are quadratic and fractional, which makes it more challenging to be solved. Although there is no general approach to solving them optimally, in the following section IV, we reformulate it and prove that the obtained solution can be approximated as the solution of the original problem.

\section{Performance Analysis and Secret Rate Maximisation}\label{sec:opt}
In this section, we address the ASR-oriented optimization problems for cases of UAV macro mobility and MA micro mobility. To ensure tractable solutions, the BCD method decomposes each problem into two subproblems. AO algorithm then iteratively updates UAV's trajectory/MA's antenna positioning and their associated BF vectors.
Finally, a detailed analysis of the computational complexity is provided to evaluate the algorithmic efficiency.

\subsection{Optimization of antenna positioning and BF matrix for MA micro-mobility} 
To maximize the ASR across all time slots, joint optimization focuses on the x-axis positions of MA elements $\mathbf{x}_{m}$ and the BF vector $\mathbf{w}$. To overcome the non-smooth objective function in Problem \eqref{eq:P1} caused by the $[\cdot]^+$ operator, which can be eliminated safely through transmission power control. Specifically, transmission suspension occurs when $\tau[n] < 0$ \cite{Sun2021Secure}.

\subsubsection{Antenna positioning optimization for given BF vector}
The MA position matrix $\mathbf{x}_{m}$ is optimized while fixing both the UAV's initial position and the BF vector $\mathbf{w}$. Since the MA's displacement range operates at centimeter-to-millimeter scales, it induces negligible variation in the departure angle under far-field propagation conditions. Consequently, the angle $\alpha_i$ remains effectively constant. Let $d_{u, i}$ be the distance between the UAV and ground user $i$. To sum up, the steering vector can be expressed as
\begin{equation} 
\mathbf{a}_i[n] = 
\begin{bmatrix} 
e^{j\frac{2\pi}{\lambda} x^{\rm Loc}_1[n] \cos\alpha_{u, i}} ,&
\cdots &,
e^{j\frac{2\pi}{\lambda} x^{\rm Loc}_M[n] \cos\alpha_{u, i}}
\end{bmatrix}^\mathbf{T}. 
\end{equation}
The MA positioning problem thus reduces to a computationally tractable form:
\begin{subequations} \label{eq:P1'}
\begin{align}
\textbf{P1-1: }
\max_{\mathbf{x}_{ m }} \quad & \frac{1}{N}\sum_{n=1}^{N} \left[ { \log_2{\left( 1+ \frac{ \beta_0 \left| \mathbf{a}_b^{\rm \mathbf{H}}[n] \mathbf{w} \right|^2 }{ d_{u, b}^2 \, \sigma_b^2 }\right)} }  \right. \notag \\
&\left.- { \log_2{\left( 1+ \frac{ \beta_0 \left| \mathbf{a}_e^{\rm \mathbf{H}}[n] \mathbf{w} \right|^2 }{ d_{u, e}^2 \, \sigma_e^2 }\right)} } \right] \\
\mbox{s.t.}\quad 
& \left(\ref{eq:P1-2s}\right), \left(\ref{eq:P1-3}\right), \left(\ref{eq:P1-4}\right).\notag
\end{align}
\end{subequations}

Problem \eqref{eq:P1'} still is a non-covex program of $ \mathbf{x}_ \mathbf{MA}$, since the dependence of the steering vector on antenna positions is expressed through complex exponential functions, with the resulting function proven to be non-convex with respect to $\mathbf{a}_i$. A PGA algorithm to compute a high-quality feasible solution solution. The gradient of $ \mathbf{\tau}[n]$ with respect to \textcolor{blue}{$\mathbf{x}_m[n]$} is given by
\begin{equation}
\begin{aligned}
\nabla_{\mathbf{x}_{m}[n]} \tau[n] = & \frac{\beta_0}{\ln 2}
\left( 
\frac{2\mathbf{a^{\rm \mathbf{H}}}_b[n] \mathbf{w}[n] \cdot \nabla_{\mathbf{x}_{m}[n]} \mathbf{a^{\rm \mathbf{H}}}_b[n] \cdot \mathbf{w}[n]}{d_{u, b}^2[n] \sigma_b^2 (1 + A[n])} \right. \\
& \left. - 
\frac{2\mathbf{a^{\rm \mathbf{H}}}_e[n] \mathbf{w}[n] \cdot \nabla_{\mathbf{x}_m[n]} \mathbf{a^{\rm \mathbf{H}}}_e[n] \cdot \mathbf{w}[n]}{d_{u, e}^2[n] \sigma_e^2 (1 + B[n])}
\right),
\end{aligned}
\end{equation}
where $A[n] = \frac{\beta_0 \left| \mathbf{a}_b^{\rm \mathbf{H}}[n] \mathbf{w}[n] \right|^2}{d_{u, b}^2[n] \sigma_b^2}$, and 
$B[n] = \frac{\beta_0 \left| \mathbf{a}_e^{\rm \mathbf{H}}[n] \mathbf{w}[n] \right|^2}{d_{u, e}^2[n] \sigma_e^2}$. 
The derivative of $ \mathbf{a}_i[n]$ with respect to $ \mathbf{x}_{\mathrm{MA}}[n] $ is given by \cite{Hu2024Secure,Wang2024Movable}
\begin{align}
\nabla_{\mathbf{x}_{\mathrm{MA}}[n]} \mathbf{a}_i[n]= 
\left[
\frac{\partial a_{i}[n]}{\partial x^{\mathrm{Loc}}_1[n]},\ 
\dots,\ 
\frac{\partial a_{i}[n]}{\partial x^{\mathrm{Loc}}_M[n]}
\right].
\label{eq:jacobian_diag}
\end{align}
Let symbol of $^*$ denote the complex conjugate. By expressing $a_i[n]^*$ in trigonometric form by using Euler’s formula, $e^{jx}=(\cos x +j \sin x)$ \cite{moskowitz2002complex}, its derivative can be characterized as
\begin{equation}
\begin{aligned}
\nabla_{x^{\rm Loc}_m[n]} a_i[n]^* =& \frac{2\pi}{\lambda} \cos \alpha_{ i} \cdot 
\left[
-\sin \left( \frac{2\pi}{\lambda} x^{\rm Loc}_m[n] \cos \alpha_{i} \right) \right. \\
& \left.
- j \cos \left( \frac{2\pi}{\lambda} x^{\rm Loc}_m[n] \cos \alpha_{ i} \right)
\right].
\end{aligned}
\end{equation}
Building on this formulation, we integrate the gradient of $\tau[n]$ into the AdaGrad optimization framework to iteratively optimize MA element positions. Specifically, according to the accumulated historical gradient information, the position of the $m$-th antenna element at $(k+1)$-th iteration is updated as
\begin{equation}
{x^{\rm Loc}_m}^{(k+1)}[n] = {x^{\rm Loc}_m}^{(k)}[n] + \zeta_{\text{ada}} \cdot \nabla_{x^{\rm Loc}_m[n]} \tau[n],
\end{equation}
which adaptively adjusts the learning rate for each dimension, and 
where $\zeta_{\text{ada}}$ represents the adaptive step size. To enforce the movement constraint (\ref{eq:P1-3}) maintaining minimum safety distances between adjacent antennas, each position update undergoes a feasibility check to verify compliance with spatial constraints. That is,
\begin{equation}
{x^{\rm Loc}_m}^{\left( k+1 \right)}[n] = {x^{\rm Loc}_m}^k[n] + d_{\min}^{\rm MA} \cdot \frac{ {x^{\rm Loc}_m}^{\left( k+1 \right)}[n] - x^{\rm Loc}_m[n-1]}{\left\| {x^{\rm Loc}_m}^{\left( k+1 \right)}[n] - x^{\rm Loc}_m[n-1] \right\|}
\end{equation}

Subsequently, the feasible antenna positioning is rigorously constrained within the predefined spatial boundaries satisfying the constraint in (\ref{eq:P1-3}), ensuring strict operational region compliance. Finally, the suboptimal antenna positioning in the $n$-th time slot can be determined by
\begin{equation}
{x^{\rm Loc}_m}^{*}[n] = \min \left( \max \left( {x^{\rm Loc}_m}^{ \left( k+1 \right)}[n], x_{\min}^{m, \rm Loc} \right), x_{\max}^{m, \rm Loc} \right).
\end{equation}

\subsubsection{BF optimization given positions of MA elements}\label{sub:BF design MA}
Given a fixed configuration of the MA element positions, the objective is to maximize the ASR by exclusively optimizing the BF vector $ \mathbf{w} $, subject to the power constraint in Eq. \eqref{eq:P1-4}. The subproblem is formulated as follows:
\begin{subequations} \label{eq:P1-2}
\begin{align}
\textbf{P1-2: }
\max_{\mathbf{w}_{\mathbf{MA}}} \quad & \frac{1}{N}\sum_{n=1}^{N} \left[ { \log_2{\left( 1+ \varepsilon_1 \left| \mathbf{a}_b^{\rm \mathbf{H}}[n] \mathbf{w}[n] \right|^2  \right)} }  \right. \notag \\
&\left.- { \log_2{\left( 1+ \varepsilon_1 \left| \mathbf{a}_e^{\rm \mathbf{H}}[n] \mathbf{w}[n] \right|^2 \right)} } \right] \\
\mbox{s.t.}\quad 
&  \left(\ref{eq:P1-4}\right).\notag
\end{align}
\end{subequations}
where $\varepsilon_1 = \frac{\beta_0}{d_{u, b}^2 \sigma_b^2}$ and $\varepsilon_2 = \frac{\beta_0}{d_{u, e}^2 \sigma_e^2}$.

For the non-convex optimization of $\mathbf{w}[n]$, PGA method is adopted again, which is consistent with the approach employed in solving Problem \eqref{eq:P1'}.
The gradient of the objective function with respect to $\mathbf{w}[n]$ is given by
\begin{equation} \label{eq:w_tau}
\begin{aligned}
\nabla_{\mathbf{w}[n]} \tau[n] =& \frac{\beta_0}{\ln 2} \left[
\frac{\varepsilon_1[n] \nabla_{\mathbf{w}[n]} \left| \mathbf{a}_b^{\rm \mathbf{H}}[n] \mathbf{w}[n] \right|^2}{1 + \varepsilon_1[n] \left| \mathbf{a}_b^{\rm \mathbf{H}}[n] \mathbf{w}[n] \right|^2} \right.\\
&\left. - \frac{\varepsilon_2[n] \nabla_{\mathbf{w}[n]} \left| \mathbf{a}_e^{\rm \mathbf{H}}[n] \mathbf{w}[n] \right|^2}{1 + \varepsilon_2[n] \left| \mathbf{a}_e^{\rm \mathbf{H}}[n] \mathbf{w}[n] \right|^2}
\right],
\end{aligned}
\end{equation}
where
\begin{align}
\nabla_{\mathbf{w}[n]} \left| \mathbf{a}_b^{\rm \mathbf{H}}[n] \mathbf{w}[n] \right|^2 &= 2 \mathbf{a}_b[n] \mathbf{a}_b^{\rm \mathbf{H}}[n] \mathbf{w}[n] ,\\
\nabla_{\mathbf{w}[n]} \left| \mathbf{a}_e^{\rm \mathbf{H}}[n] \mathbf{w}[n] \right|^2 &= 2 \mathbf{a}_e[n] \mathbf{a}_e^{\rm \mathbf{H}}[n] \mathbf{w}[n].
\end{align}
Similarly, the BF vector is updated using AdaGrad framework \cite{Kolodziej2021,Lu2021Joint} as follows
\begin{equation}
\mathbf{w}^{(k+1)}[n] = \mathbf{w}^{(k)}[n] + \delta_{\text{adag}} \nabla_{\mathbf{w}[n]} C(\mathbf{w}[n]).
\end{equation}
A projection is applied to ensure the updated vector satisfies the power constraint
\begin{equation} \label{eq:w(k+1)}
\mathbf{w}^{(k+1)*}[n] = \Pi \left\{ \mathbf{w}^{(k+1)}[n], \; \mathbf{w}'[n] \right\},
\end{equation}
where the normalized vector $\mathbf{w}'[n]$ is defined by \cite{10887653,Yujia2024}
\begin{equation} \label{eq:w}
\mathbf{w}'[n] = \mathbf{w}[n] \sqrt{ \frac{P_{\max}}{q} },
\end{equation}
where $q = \text{tr}(\mathbf{w}[n] \mathbf{w}^{\rm H}[n])$ denotes the instantaneous transmit power associated with the unnormalized BF vector.

\subsubsection{Simulated annealing algorithm for convergence}
To mitigate local optimal convergence risks, we employ the simulated annealing (SA) method \cite{Nguyen2021}, a well-established stochastic optimization technique. Candidate solutions are probabilistically accepted with probability $p$, which is defined as 
\begin{equation} \label{eq:p}
p = 
\begin{cases}
1, & \tau[n]^{(k+1)} > \tau[n]^{(k)} \\
\exp\left( \dfrac{\tau[n]^{(k+1)} - \tau[n]^{(k)}}{T^{(k+1)}} \right), & \tau[n]^{(k+1)} \leq \tau[n]^{(k)}
\end{cases}
\end{equation}
where $T^{(k+1)} = \rho T^{\left( k \right)}$ denotes the temperature at the $(k+1)$-th iteration, and $\rho$ is the cooling factor. By avoiding premature convergence to local optima in antenna positioning and BF of the MA obtained by PGA, the simulated annealing method enables the optimization process to explore a broader solution space. Consequently, a near-optimal solution for Problem \eqref{eq:P1} is attained. Although this solution may not represent the global optimum, it delivers a satisfactory high-quality outcome.

\subsection{UAV-based macro-mobility optimization} 

By jointly optimizing UAV trajectory $\mathbf{q_{UAV}}$ and BF
$\mathbf{w}$, we aim to maximize the ASR over all time slots. Similar to Problem \eqref{eq:P1}, the $[\cdot]^+$ operator can be also omitted safely. 

\subsubsection{UAV trajectory optimization for given BF vector}
The UAV trajectory is optimized for a predefined BF vector with ULA inter-element spacing fixed at $d^{\rm ULA}_{\text{fix}}$. The steering vector is
\begin{equation}
\mathbf{a}_i[n] = 
\begin{bmatrix} 
e^{\frac{j 2 \pi}{\lambda}  x^{\rm Loc}_{1} \cos \alpha_{i}[n]}, & 
\ldots ,& 
e^{j\frac{2 \pi}{\lambda} x^{\rm Loc}_M \cos \alpha_{i}[n]}
\end{bmatrix}^\mathbf{T},
\end{equation}
where $ x^{\rm Loc}_m-x^{\rm Loc}_{m-1}= d^{\rm FPA}_{\text{fix}}$ for $m \in \{2,...,M\}$.

Crucially, Eq. \eqref{eq:SNR_b} and Eq. \eqref{eq:SNR_e} show that the distance term $d_{u, i}$ exhibits higher sensitivity to $\mathbf{q}_{\text{UAV}}$ than the steering vector $\mathbf{a}_i$. We approximate $\mathbf{a}_i[n]$ at the $(k+1)$-th iteration  using that of the $k$-th iteration trajectory $\mathbf{q}_u[n]$, since successive trajectory updates yield minimally displaced UAV positions \cite{Liu2025UAVMA}. Thus, corresponding Subproblem can be formulated as:
\begin{subequations} \label{eq:P2-1}
\begin{align}
\text{P(2-1):} \quad 
&\max_{\mathbf{q_{u}}} \ \frac{1}{N} \sum_{n=1}^{N} 
\left[
\log_2 \left( 1 + \frac{s_1[n]}{d_{u, b}^2[n]} \right)  \right. \\
& \left.- 
\log_2 \left( 1 + \frac{s_2[n]}{d_{u, e}^2[n]} \right)
\right]\\
\text{s.t.} \quad& \left( \ref{eq:P2-2}\right), \left( \ref{eq:P2-3}\right).\notag
\end{align}
\end{subequations}
where $s_1[n]= \frac{\beta_0 \left| \mathbf{a}_b^{\rm \mathbf{H}}[n] \mathbf{w}[n] \right|^2}{\sigma_b^2}$   and   $s_2[n]= \frac{\beta_0 \left| \mathbf{a}_e^{\rm \mathbf{H}}[n] \mathbf{w}[n] \right|^2}{\sigma_e^2}$.
Under this configuration, the UAV’s positional variation primarily impacts the distance $d_{u, i}$. Nevertheless, the sub-problem is a non-convex integer programming problem that is difficult to be optimally solved efficiently. To address the problem, the slack variables $u[n]$ and $w[n]$ are introduced to relax constraint Eq. (\ref{eq:2-1-1})-(\ref{eq:2-1-3}). The ASR maximization problem at the $n$-th time slot can be reformulated as
\begin{subequations} \label{eq:P2-1'}
\begin{align}
\text{P(2-1)'} \quad \max_{\mathbf{q}[n]} \quad & \left[ \log_2\left(1 + \frac{s_1[n]}{u[n]}\right) - \log_2\left(1 + \frac{s_2[n]}{w[n]}\right)\right]\\
\text{s.t.} \quad & w[n] - x_u^2[n] + 2x_e x_u[n] - x_e^2 - y_u^2[n] \notag\\
&+ 2y_e y_u[n] - y_e^2 - H^2 \leq 0, \quad \forall n\label{eq:2-1-1} \\
& x_u^2[n] - 2x_b x_u[n] + x_b^2 + y_u^2[n] - 2y_b y_u[n] \notag\\
&+ y_b^2 + H^2 - u[n] \leq 0, \quad \forall n \label{eq:2-1-2} \\
& w[n] \geq H^2 \label{eq:2-1-3} \\
& \left(\ref{eq:P2-2} \right), \left(\ref{eq:P2-3} \right) 
\end{align}
\end{subequations}
Building upon the optimization structure established in prior analysis, we initialize a feasible UAV trajectory configuration $\mathbf{q}_0 \triangleq [\mathbf{q}_0[1], \dots, \mathbf{q}_0[N]]^\mathbf{T}$,  along with the corresponding slack variable vector $\mathbf{u}_0 \triangleq [u_0[1], \dots, u_0[N]]^\mathbf{T}$. These initializations serve as the starting point for the iterative trajectory refinement procedure, ensuring that each iteration remains within the feasible region defined by the kinematic and spatial constraints. 
To approximate the non-convex objective and constraints, the logarithmic and squared terms are relaxed via First-order Taylor expansion around the initial points, and then the final optimization problem exhibits a concave objective function and convex constraints, which can be directly solved using the CVX toolbox in MATLAB \cite{He2023full,Sun2021Unmanned}. Readers are referred to \cite{He2023full,Sun2021Unmanned,wang2020completion} for comprehensive details.

\subsubsection{BF optimization for given UAV macro-mobility trajectory}
The BF vector optimization in this case can be achieved by using similar methods in the Subsection \ref{sub:BF design MA}.
.

\subsection{AO algorithm for ASR maximization}
By alternately solving the above subproblems (UAV trajectory/antenna positioning and the BF), the ASR maximization for \emph{MA micro-mobility} and \emph{UAV macro-mobility} can be efficiently achieved. First of all, SCA method is employed to iteratively approximate the optimal UAV trajectory within a stochastic programming framework. Furthermore, to mitigate gradient estimation errors induced by random variables, a Monte Carlo sampling mechanism is integrated into PGA architecture. The pesudo-codes of the PGA-based BF optimization and AO procedure are given in Algorithm \ref{PGA} and  Algorithm \ref{AO1}, respectively.

\begin{algorithm}[H] 
	\caption{PGA-based BF Optimization}  
	\begin{algorithmic}[1]  
		\State  \textbf{Initialization:} the number of antennas $M$, initial BF $\mathbf{w}^{(0)}$, the  maximum iteration number $I_{\max}$, the maximum Monte Carlo simulation number $M_t$, the gradient value $G$\\
		\textbf{Repeat}\\
		~~~~Updating $k=k+1$\\
		~~~~Updating $G=0$\\
		~~~~\textbf{Repeat}\\
		~~~~~~~~Calculating $\nabla_{\mathbf{w}}\tau$ with Eq. (\ref{eq:w_tau})\\
		~~~~~~~~Updating $G=G+\nabla_{\mathbf{w}}\tau$ \\
		~~~~\textbf{Until}  $M_t$ is reached\\
		~~~~Updating $G=G/M_t$ \\
		~~~~Calculating $\textbf{w}^{(k+1)}$ with Eq. (\ref{eq:w(k+1)})\\
		~~~~Feasibility verification of $\textbf{w}^{(k+1)}$ \\
		\textbf{Until} the maximum iteration number $I_{\max}$ is reached
	\end{algorithmic}
	\label{PGA}
\end{algorithm}

\subsubsection{Complexity analysis of Problem \eqref{eq:P1}} 
The original optimization problem \eqref{eq:P1} is decoupled into two interdependent subproblems via BCD method: (1) antenna positioning optimization using PGA, and (2) BF design via PGA. We solve these subproblems alternately to achieve progressive optimization.
The computational complexity of the BF optimization subproblem is $\mathcal{O}(I_{\max}M_t)$, dominated by the Monte Carlo sample size $M_t$ and maximum iterations $I_{\max}$. The antenna positioning optimization subproblem exhibits higher complexity $\mathcal{O}(MI_{\max}M_t)$ due to sequential position exploration for each of the $M$ antenna elements. With $I_{\max}$ AO iterations in the outer loop, the overall complexity is $\mathcal{O}\left(I_{\max}^2 M M_t\right)$.

\subsubsection{Complexity analysis of Problem \eqref{eq:P2}} 
The problem \eqref{eq:P2} can be decomposed into two subproblems: 1) the UAV trajectory optimization using the CVX toolbox; 2) BF vector optimization. The computational complexity of the BF design subproblem aligns with that of Problem \eqref{eq:P1}. In contrast, the UAV trajectory optimization subproblem is formulated as a convex optimization one and solved via CVX empowered by interior-point method. The complexity of this subproblem is quantified as $O(N^{3}_{\text{var}}\log(\epsilon^{-1})$, where $N_{\text{var}}$ denotes the number of variables, and $\epsilon$ denotes the solution accuracy. Consequently, the complete complexity is $\mathcal{O}\left( I_{\rm{iter}} ( N^{3}_{\text{var}}\log(\epsilon^{-1})   +   I_{\max}M_t)  \right)$.

\begin{algorithm}[H] 
	\caption{Proposed AO for Problems \eqref{eq:P1} and \eqref{eq:P2}}  
	\begin{algorithmic}[1]  
		\State  \textbf{Initialization:} the  maximum iteration number $I_{\max}$, initial and terminal position of UAV $q_0$ and $q_t$, the initial positions of MA $\mathbf{ x}_m^{(0)}$ (or the initial trajectory of UAV $\mathbf{q}_{u}^{(0)}$), the BF $\mathbf{ w}^{(0)}$,  $\tau_0 = f_{\mathbf{P1}}(\mathbf{ x}_{m}^{(0)}, \mathbf{ w}^{(0)})$ ($\tau_0 = f_{\mathbf{P2}}( \mathbf{q}_u^{(0)}, \mathbf{ w}^{(0)})$), index of initial iteration $i=0$, cooling factor $\rho$, and initial temperature $T^0$\\
		\textbf{Repeat}\\
		~~~~$k = k + 1$\\
		~~~~Update ${\mathbf{x}_\mathbf{MA}^{(k)}}$ (or ${\mathbf{q}_u^{(k)}}$) with initial point $\left({\mathbf{x}_m^{(k-1)}}, \mathbf{w}^{(k-1)} \right)$ (or $\left({\mathbf{q}_u^{(k-1)}}, \mathbf{w}^{(k-1)} \right)$) with Eq. \eqref{eq:P1-2} (or Eq. \eqref{eq:P2-1'} )\\
		~~~~Updating $\mathbf{w}^{(k)}$ with initial point $\left({{\mathbf{x}_m^{(k)}}_{\rm com}}, \mathbf{w}^{(k-1)}\right)$ with Eq. $\eqref{eq:w}$\\
		~~~~$\tau^{(k)}=f_{\mathbf{P1}}\left({\mathbf{x}_m^{(k)}}, \mathbf{w}^{(k)} \right)$ (or $\tau^{(k)}=f_{\mathbf{P2}}\left({\mathbf{q}_u^{(k)}}, \mathbf{w}^{(k)} \right)$).\\
		~~~~Calculating $p$ with Eq. (\ref{eq:p})\\
		
		~~~~\textbf{if} $p <  \mathbf{random}(0,1)$  \\        ~~~~~~~~$\tau^{(k)}=\tau^{(k-1)}$, $\mathbf{w}^{(k)}=\mathbf{w}^{(k-1)}$, $\mathbf{ x}_m^{(k)}= \mathbf{ x}_m^{(k-1)}$\\
		~~~~\textbf{end if} \\
		~~~~$T^{(k+1)}=\rho T^{(k)}$\\
		\textbf{Until} the maximum iteration number $I_{\max}$ is reached
	\end{algorithmic}
	\label{AO1}
\end{algorithm}

\emph{Findings:} Although both Problem \eqref{eq:P1} and Problem \eqref{eq:P2} follow a similar alternating optimization structure, the latter has higher computational complexity compared to the former due to the complexity of the trajectory optimization. In Problem \eqref{eq:P1},  the antenna positions are updated independently using a simple first-order method. By contrast, Problem \eqref{eq:P2} involves joint optimization of UAV's trajectory over multiple time slots using an interior-point method featured by second-order, resulting in higher computational complexity. Consequently, the overall complexity of Problem \eqref{eq:P2} increases substantially, especially when the number of time slots is large.

\section{Evaluations}\label{sec:simulation}
In this section, numerical simulations are presented to evaluate the secrecy performance of micro-mobility and macro-mobility and quantify their performance differences. Under a representative single-Bob single-Eve scenario, two schemes are simulated: (1) MA micro-mobility: The UAV maintains its initial position while dynamically adjusting each of $M$ antenna elements along the x-axis within a $4\lambda$ range, with BF vector jointly optimized. (2) UAV macro-mobility: The UAV equipped with a $4\lambda\cdot M$-spaced ULA flies at a fixed altitude of $H$, with trajectory and BF optimized for secrecy performance.
Key simulation parameters are listed in Table \ref{tab:sim para validation}.

\begin{table}[!htb]  \caption{Simulated parameters and values}
\centering
\label{tab:sim para validation}
\begin{tabular}{lll}
\toprule
  Symbol & Meanings &Values\\
\midrule
  $H$          & the flight altitude of UAV     & $\{50, 100\}$m \\
  $q_b$        & the position of Bob            & $(0,0,0)$ \\
  $q_e$        & the position  of Eve          & $(400,0,0)$ \\
  $q_0$     & the initial position of UAV   & $(200,200,H)$ \\
  $q_t$     & the target position of UAV      & $(200,-200,H)$ \\
  $\alpha$     & path-loss exponent        & 2\textsuperscript{\cite{Hu2025Movable?}}        \\
  $M$          & the number of antennas    & $[2,8]$\textsuperscript{\cite{Tominaga2025On}} \\
  $N$          & number of time slots    & $[20,60]$\cite{Yu2025SuRLLC} \\
  $F$          &frequency band of communication & 28GHz\textsuperscript{\cite{Yujia2024}}\\
  $\lambda$    & wavelength                &0.0107m\textsuperscript{\cite{Yujia2024}}    \\
  $T$          & Initial temperature  & 1  \\
  $\rho$      &cooling coefficient & 0.8\textsuperscript{\cite{Yujia2024}}\\
  $P$         & communication power        &   $[0.1, 10]$W  \\
  $a_0^{\rm{max}}$    & maximum acceleration of the UAV               & 3$\rm{\mathrm{m/s^2}}$ \\
  $v_{\rm{max}}$   & maximum speed of the UAV    & 15m/s \\  
    $d_{\min}^{\rm MA}$         & Minimum inter-MA distance   & $1/2\lambda$\textsuperscript{\cite{Zuo2025}} \\ 
  \bottomrule
\end{tabular}
\end{table}

\subsection{Convergence behavior analysis}
\begin{figure}[!ht]
\centering
\includegraphics[width=2.7in]{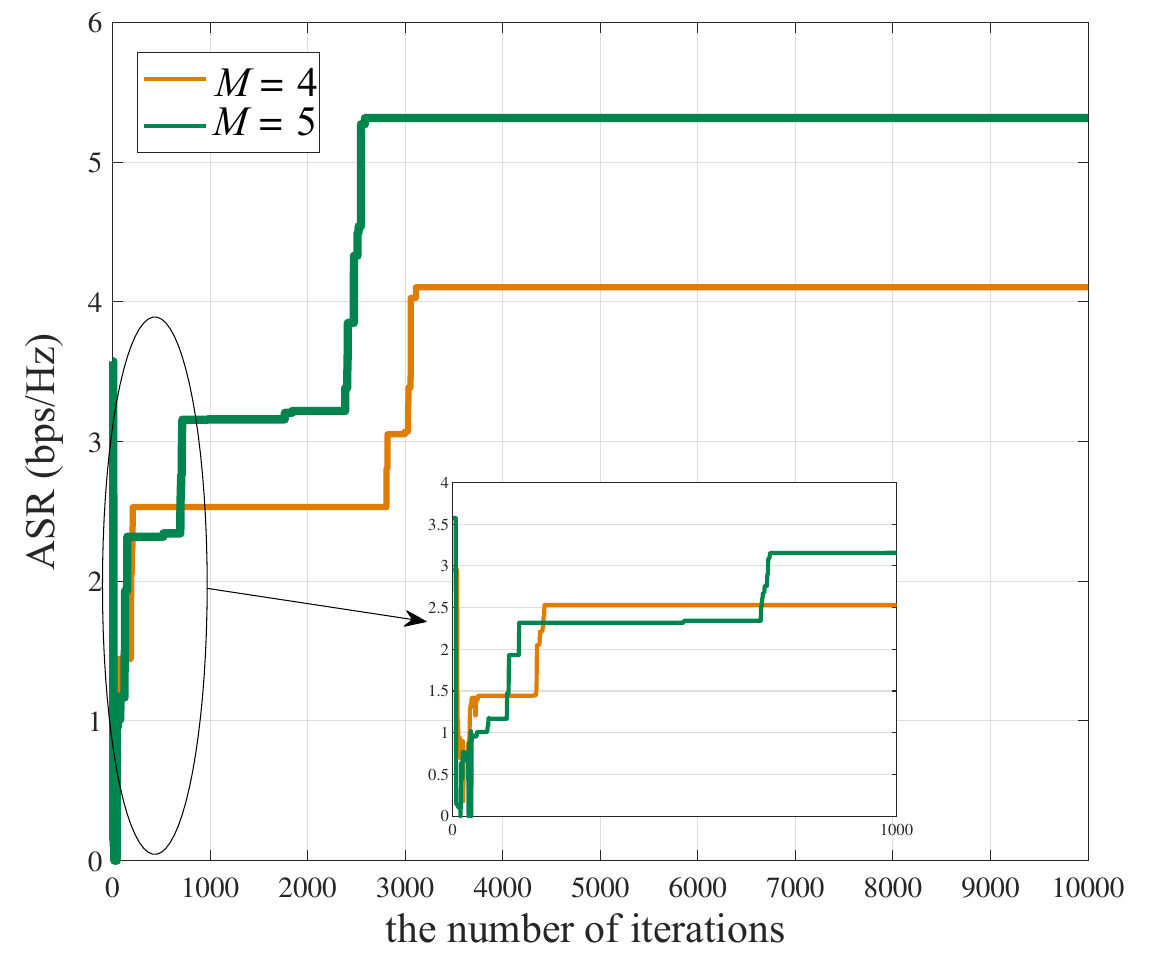}
\caption{\small ASR of MA micro-mobility \emph{vs.} the number of iterations}
\label{MA_SR_p=1_(iter)}
\end{figure}

\begin{figure}[!ht]
 	\centering
 	\includegraphics[width=3.2in]{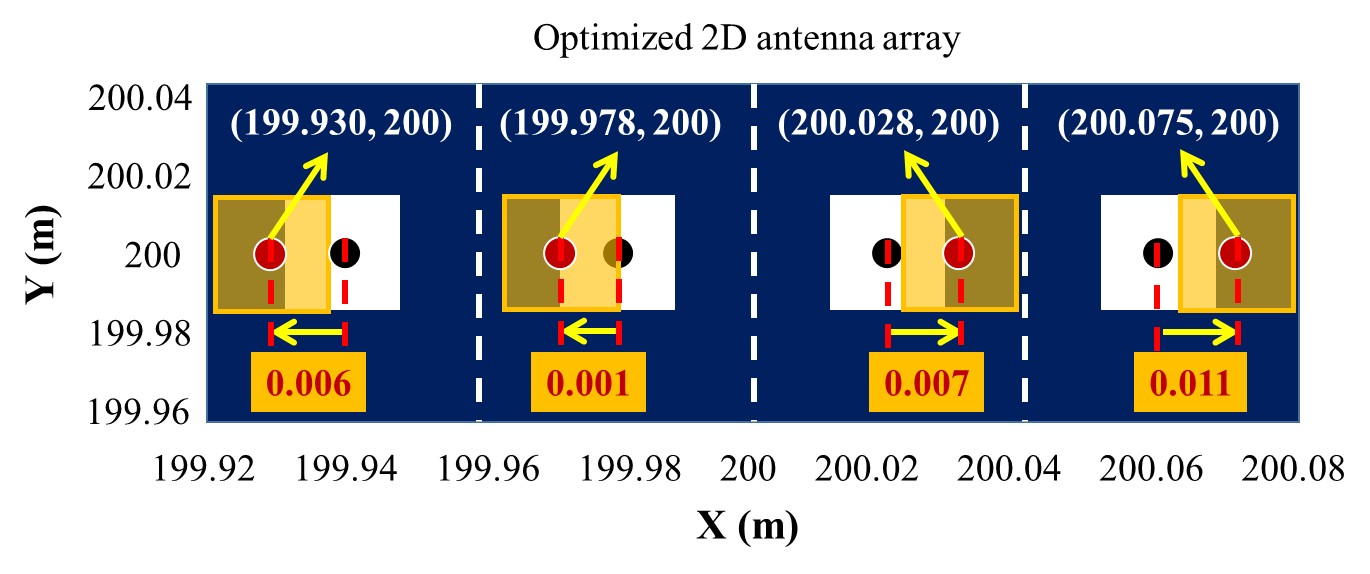}
   \caption{Optimal antenna positions ($M=4$)} \label{fig:antenna_positions_4}
 \end{figure}

\begin{figure}[!ht]
 	\centering
 	\includegraphics[width=3.2in]{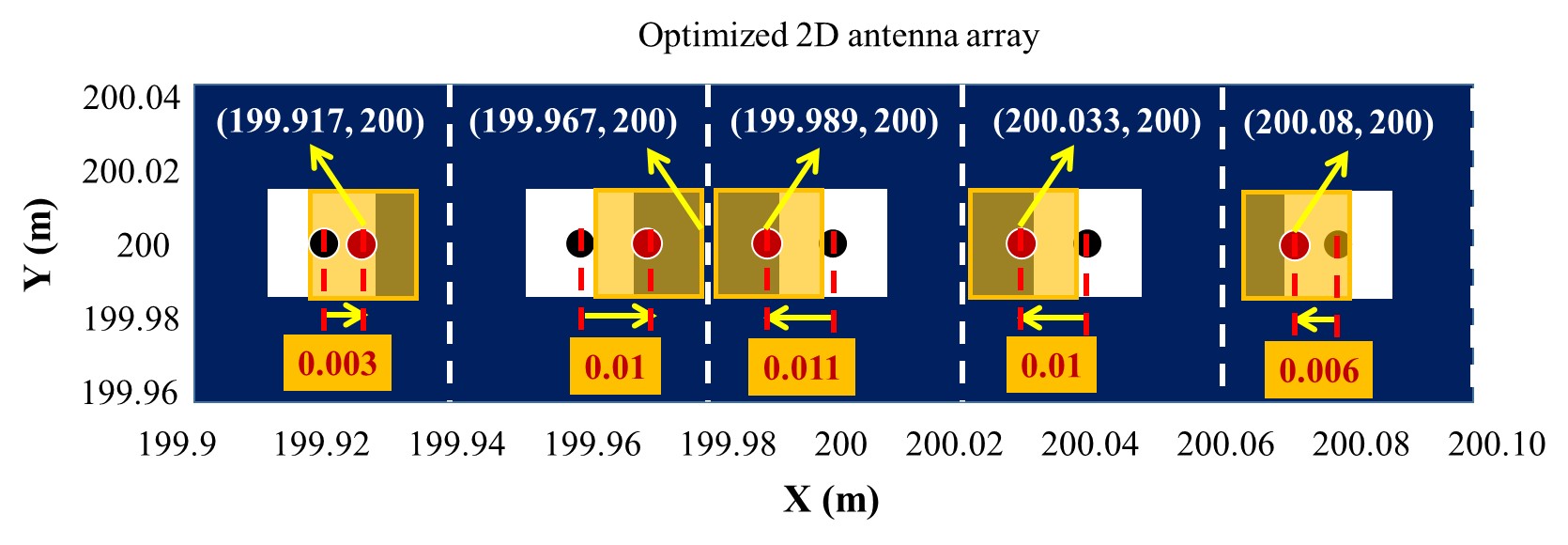}
 	\caption{Optimal antenna positions ($M=5$)} \label{fig:antenna_positions_5}
 \end{figure}

The convergence performance of the proposed AO algorithm, optimizing MA positioning, is validated under the setting of $P$=1W and $H$=50m.
Fig. \ref{MA_SR_p=1_(iter)} shows the ASR versus the number of iterations for the MA micro-mobility. Significant ASR fluctuations occur within the first 1,000 iterations for $M$=4, attributable to the probabilistic acceptance mechanism in the SA algorithm that prevents local optimum trapping. Beyond this point, the ASR stabilizes progressively, converging at th 3207-th iteration to a steady-state value of 4.1046bps/Hz. For $M$=5, pronounced oscillations persist until 2595 iterations before converging to 5.315bps/Hz. The stochastic nature of SA causes minor convergence-timing variations across trials, though all instances converge reliably within the 10000-iteration threshold. The ASR enhancement with the number of antenna elements, stems from improved spatial diversity and BF resolution, which will be  analyzed in the following Subsection. Optimized MA configurations for $M$=4 and $M$=5 are shown in Fig. \ref{fig:antenna_positions_4} and Fig. \ref{fig:antenna_positions_5}, respectively.

\begin{figure}[!ht]
\centering
\includegraphics[width=2.7in]{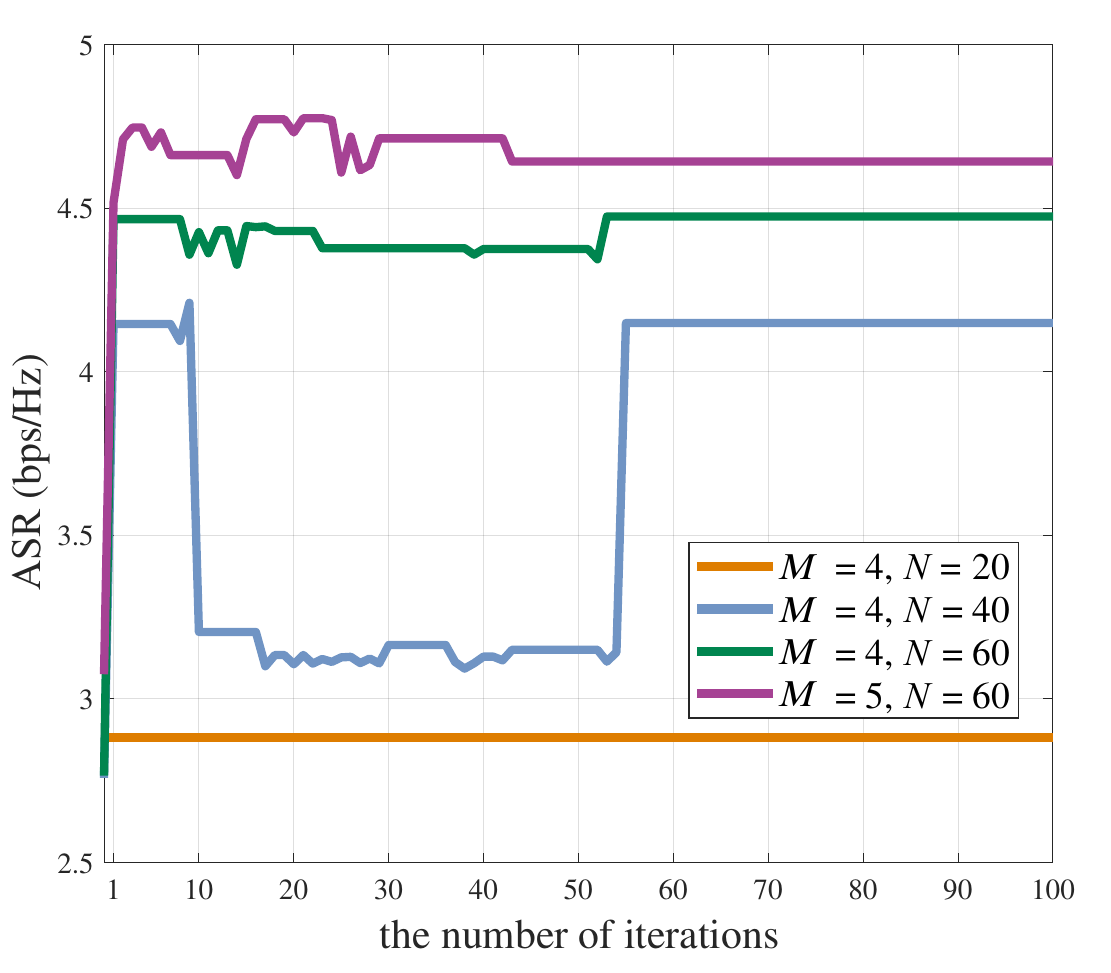}
\caption{\small ASR of UAV macro-mobility \emph{vs.} the number of iterations}
\label{fig:UAV_SR_p=1_(M_T)}
\end{figure}

\begin{figure}[!ht]
\centering
\includegraphics[width=2.7in]{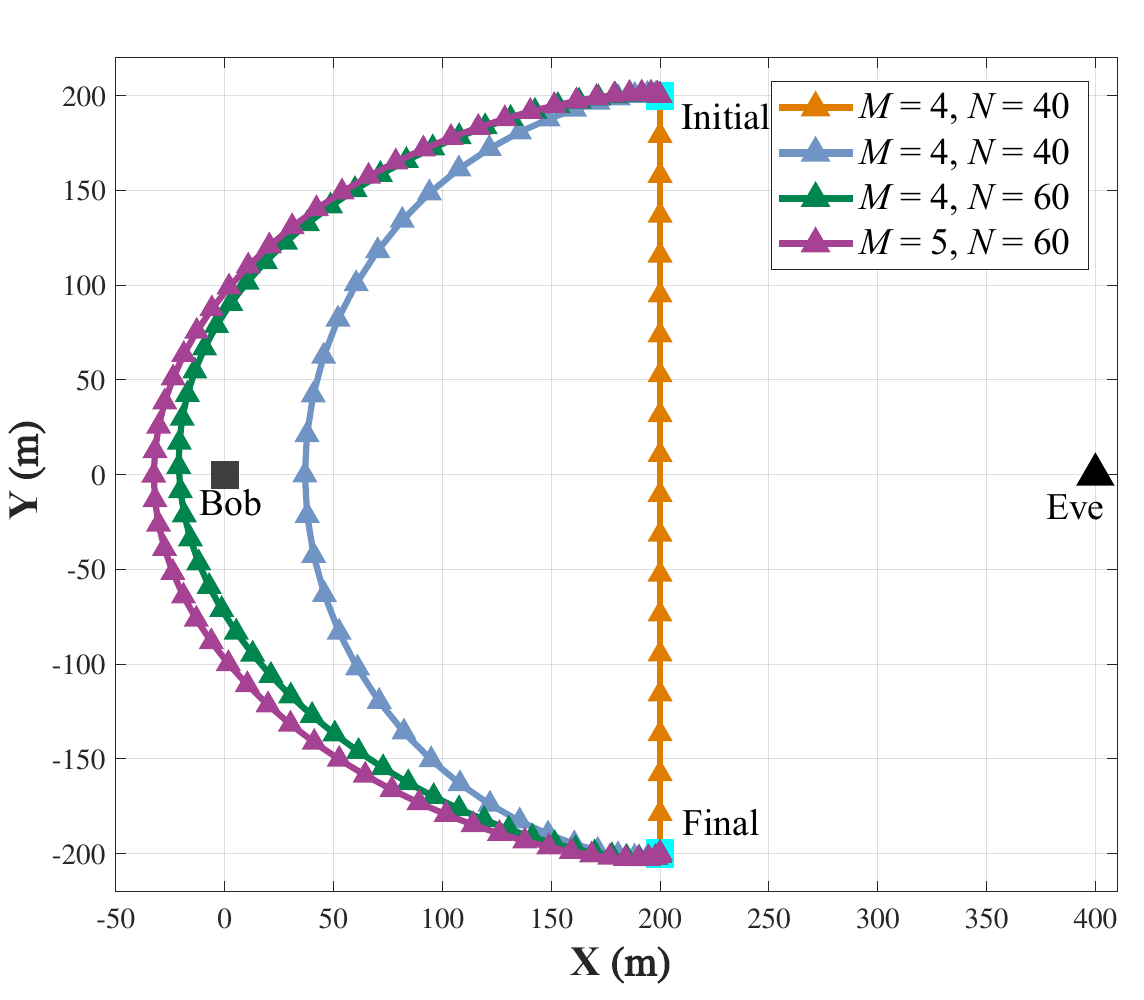}
\caption{\small Flight trajectory of UAV}
\label{fig:UAV_pos_p=1_(M_T)}
\end{figure}

Fig. \ref{fig:UAV_SR_p=1_(M_T)} illustrates convergence behavior and ASR variation for the UAV macro-mobility. The effective optimization steps total approximately 6,000 (100 outer-loop iterations$\times$60 internal CVX optimizations). Despite initial fluctuations, ASR stabilizes within 100 iterations, demonstrating reliable convergence. Performance improves progressively with longer time slots, as extended operational time enables more flexible trajectories. Fig. \ref{fig:UAV_pos_p=1_(M_T)} confirms the UAV dynamically approaches Bob while distancing from Eve, reducing interception risk. A longer flight duration enables a more flexible trajectory, allowing the UAV to explore more favorable spatial configurations. However, when the flight duration is less than $N = 20$, the UAV is constrained to follow an almost straight-line trajectory toward its destination, limiting its ability to optimize secrecy performance.

\subsection{ASRs for MA micro-mobility and UAV macro-mobility}
\begin{figure}[!ht]
\centering
\includegraphics[width=2.7in]{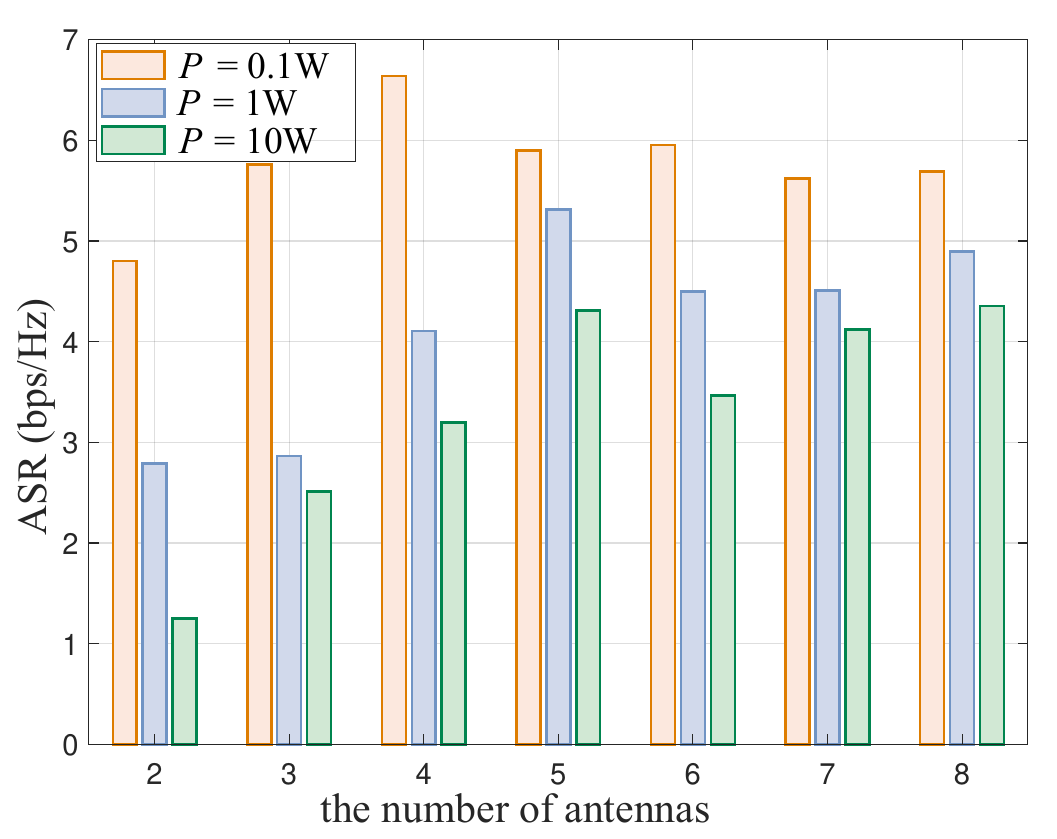}
\caption{\small ASR of MA micro-mobility \emph{vs.} the number of antennas}
\label{fig:MA-altitude50}
\end{figure}

\begin{figure}[!ht]
\centering
\includegraphics[width=2.7in]{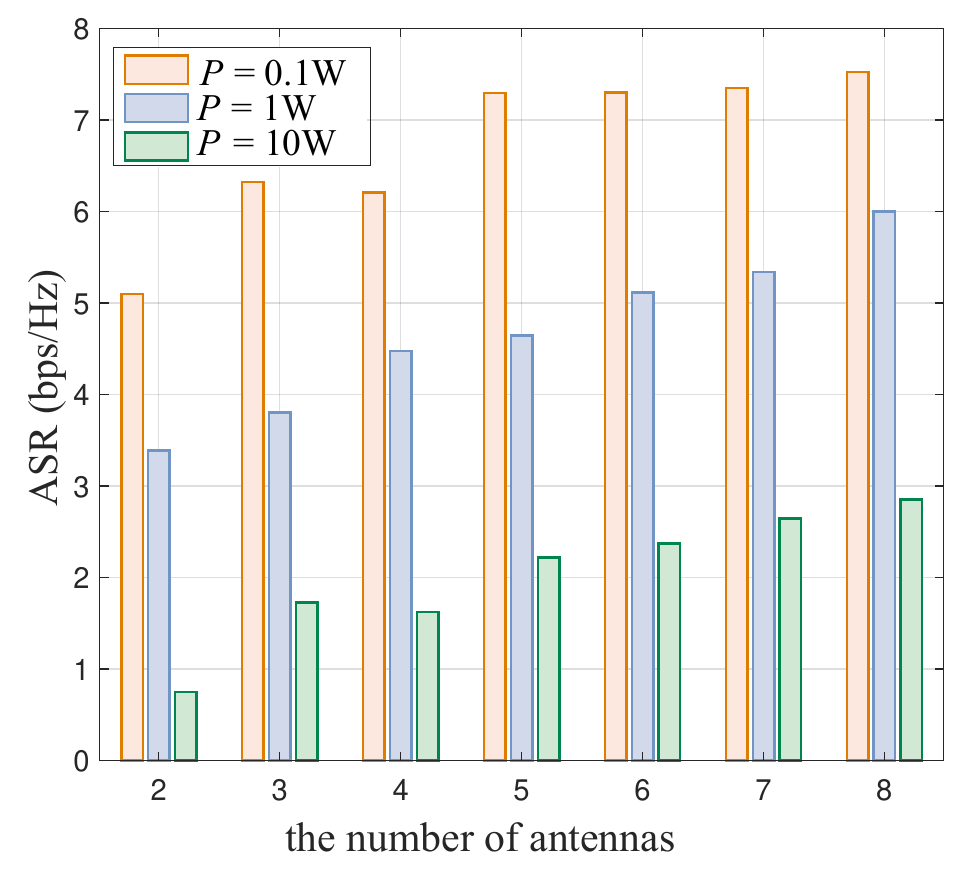}
\caption{\small ASR of UAV macro-mobility \emph{vs.} the number of antennas}
\label{fig:UAV-altitude50}
\end{figure}

\begin{figure}[!ht]
\centering
\includegraphics[width=2.7in]{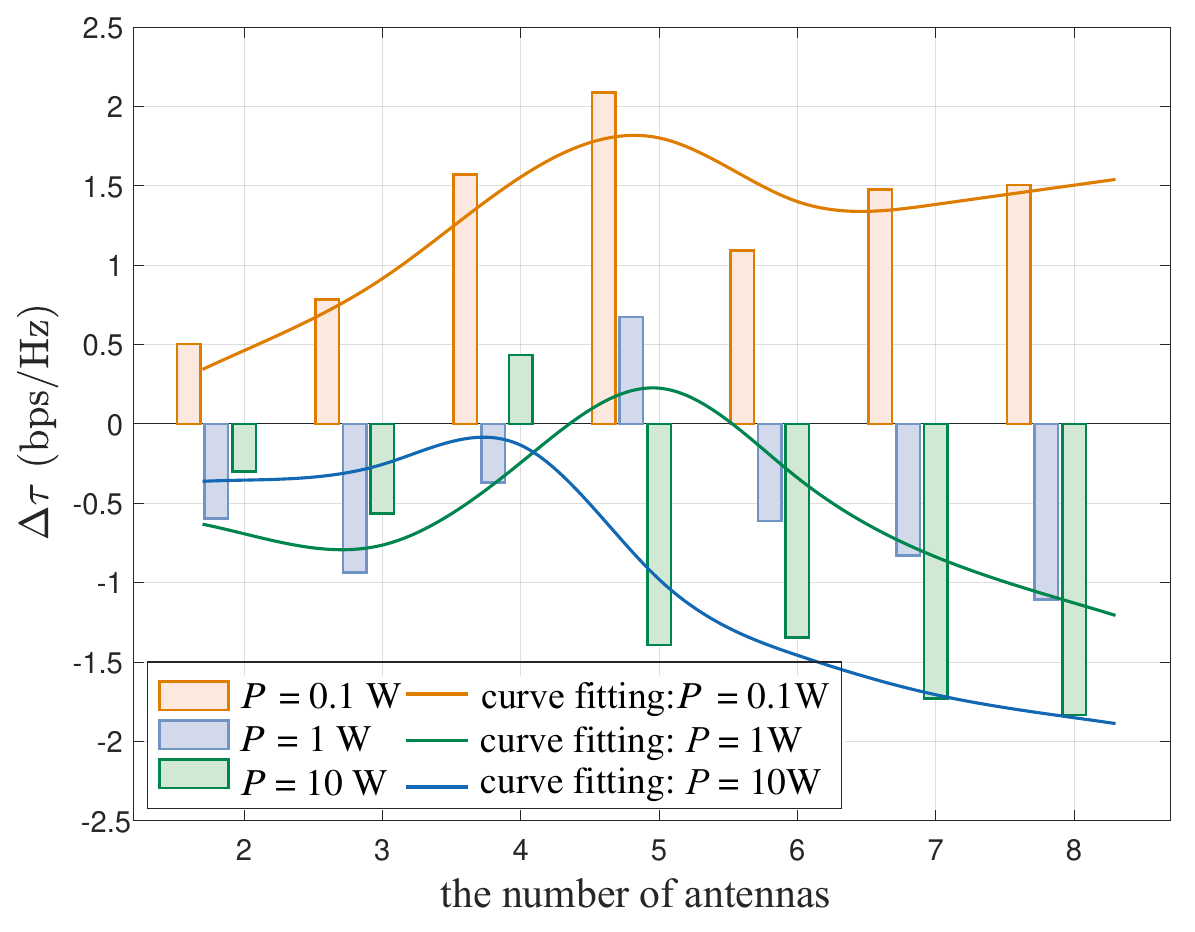}
\caption{\small $\Delta\tau$ \emph{vs.} the number of antennas}
\label{fig:MA-UAV-altitude50}
\end{figure}

The MA micro-mobility and UAV macro-mobility are systematically evaluated under different settings of antenna counts and transmission powers, with comparative analysis across two mobility paradigms at $H=$50m.
Fig. \ref{fig:MA-altitude50} shows the ASR versus the number of antennas for the MA micro-mobility. It can be noticed that ASR increases with the number of antennas when $M<5$ due to enhanced spatial degrees of freedom, but declines over the number of antennas increasing from spatial constraints and mutual coupling effects. Conversely, Fig. \ref{fig:UAV-altitude50} demonstrates ASR improvement with $M$ for the UAV macro-mobility. Both schemes exhibit significant ASR gains with higher transmission powers (0.1W$\rightarrow$10W), attributed to improved signal fidelity and SNR of legitimate user.
Fig. \ref{fig:MA-UAV-altitude50} quantifies the performance gap: The MA micro-mobility outperforms at 0.1W, while the UAV macro-mobility dominates at higher power ($P\geq1$W) except under specific conditions. Given that the solutions derived from SA-based algorithms are suboptimal and exhibit slight variations with each execution, fitting curves are utilized to effectively capture the overall trend, providing a clear and intuitive representation of the system's performance when plotted for distinct values of $M$. Notably, performance difference degradation at $M=\{4,5\}$ in MA micro-mobility confirms excessive antenna movement provides diminishing returns.

\begin{figure}[!ht]
\centering
\includegraphics[width=2.7in]{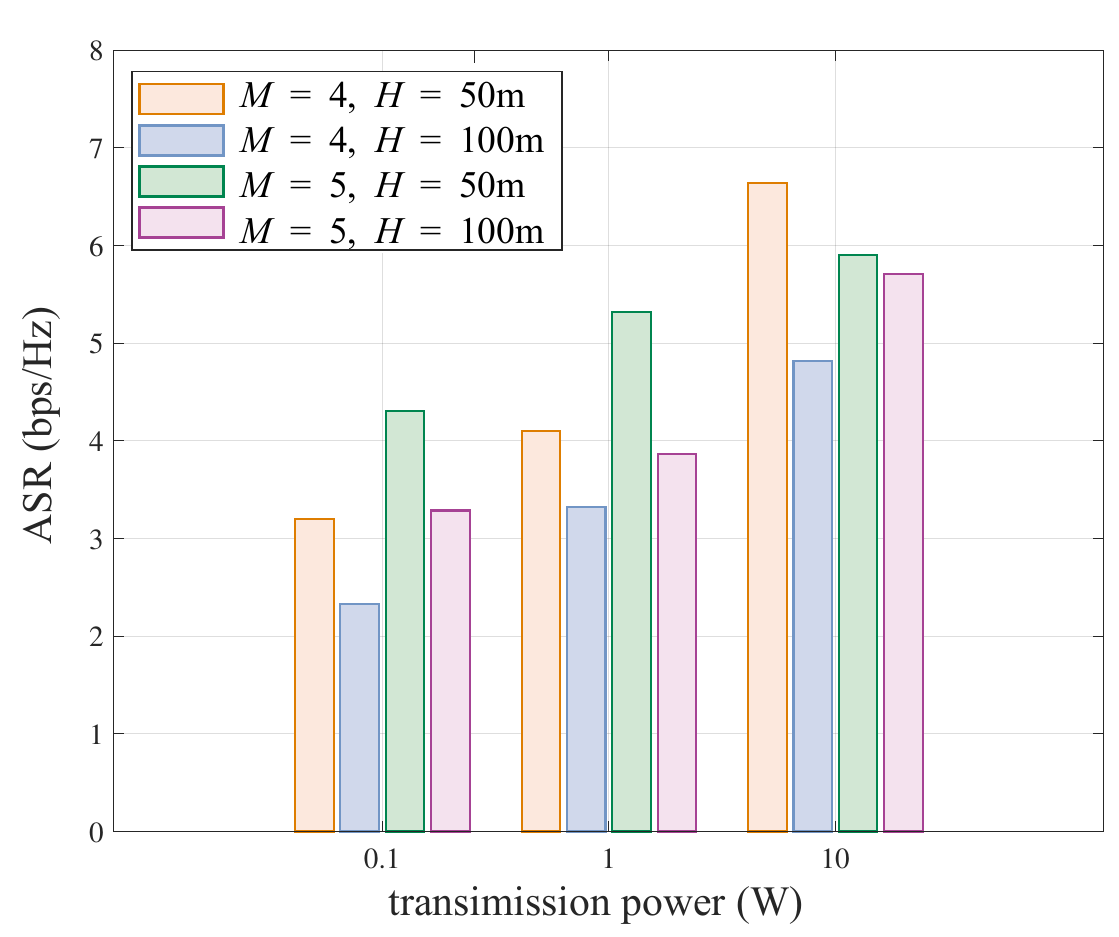}
\caption{\small ASR of MA micro-mobility \emph{vs.} transmission power}
\label{fig:MA_P_H}
\end{figure}

\begin{figure}[!ht]
\centering
\includegraphics[width=2.7in]{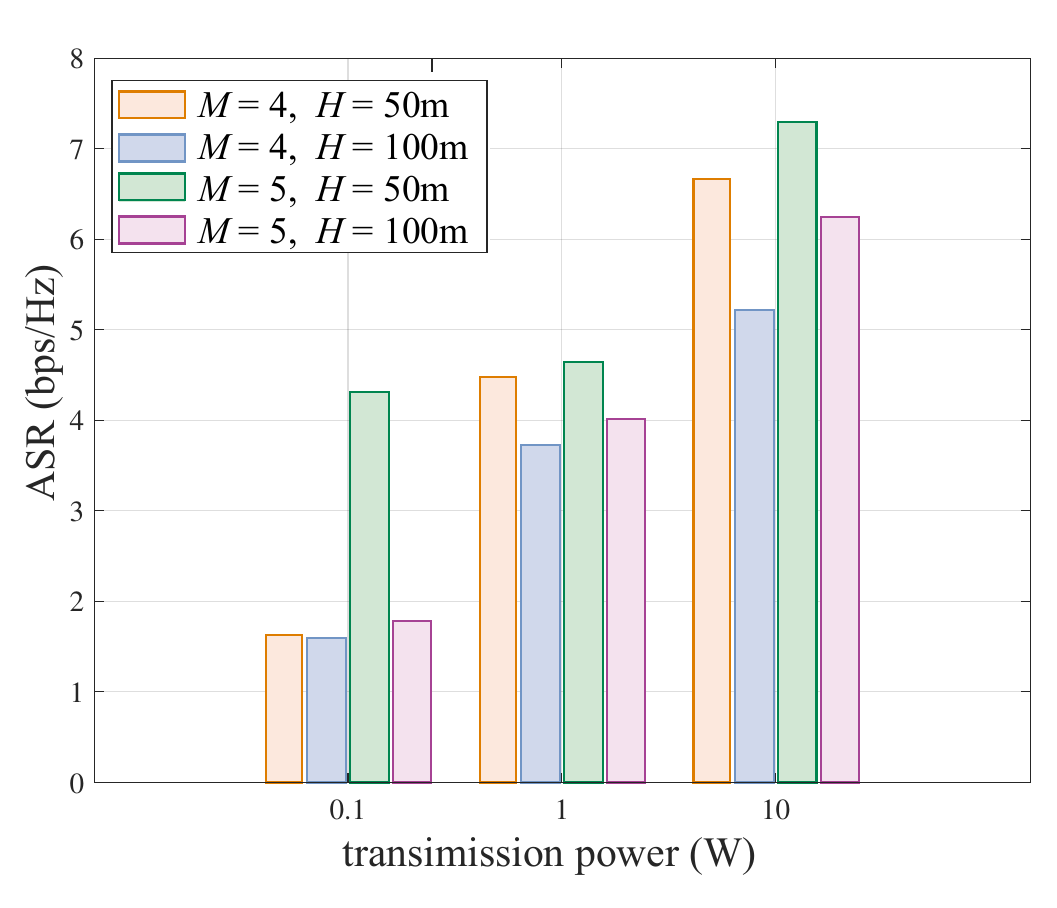}
\caption{\small ASR of UAV macro-mobility \emph{vs.} transmission power}
\label{fig:UAV_P_H}
\end{figure}

\begin{figure}[!ht]
\centering
\includegraphics[width=2.7in]{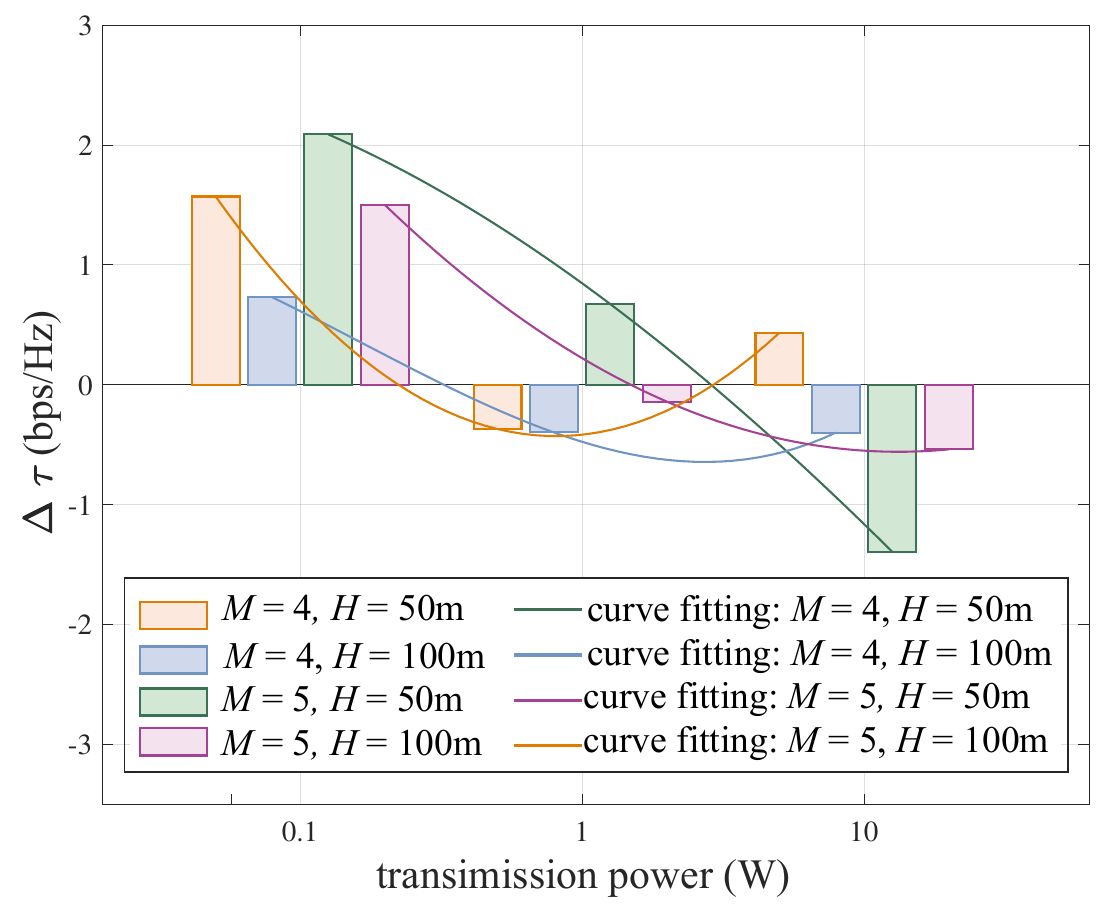}
\caption{\small $\Delta\tau$ \emph{vs.} transmission power}
\label{fig:MA-UAV_P_H}
\end{figure}

The impacts of transmission power and flight altitude on the ASR achieved by MA micro-mobility and UAV macro-mobility are illustrated in Fig.~\ref{fig:MA_P_H} and Fig. \ref{fig:UAV_P_H}. We can confirm significant ASR improvements with increasing transmission power across both mobility paradigms, consistent with prior observations in Fig. \ref{fig:MA-altitude50} and \ref{fig:UAV-altitude50}.
Performance at $H$=50m consistently surpasses that of $H$=100m due to reduced path loss to legitimate users, enhanced path loss disparity between legitimate and eavesdropper channels, and improved geometric anti-interception advantage. Fig. \ref{fig:MA-UAV_P_H} quantifies the cross-scheme performance gap ($\Delta\tau$), revealing MA-micro-mobility superiority at low power ($P$=0.1W), while UAV-macro-mobility dominates at higher power ($P$$\geq$1W) and altitudes owing to superior spatial reconfiguration capability – particularly evident in the performance saturation of MA micro-mobility at $M$=5 and $H$=50m versus the sustained gains in UAV configurations.

\begin{figure}[!ht]
\centering
\includegraphics[width=2.7in]{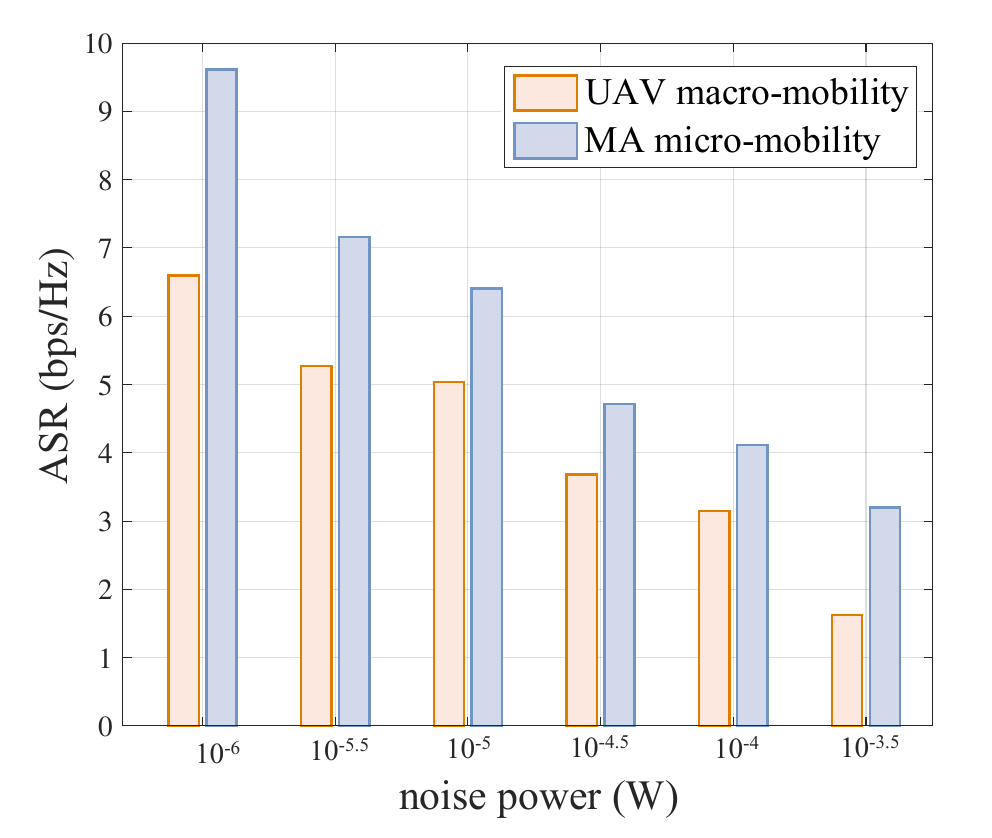}
\caption{\small ASR \emph{vs.} noise power}
\label{fig:MA_and_UAV_noise}
\end{figure}

\begin{figure}[!ht]
\centering
\includegraphics[width=2.7in]{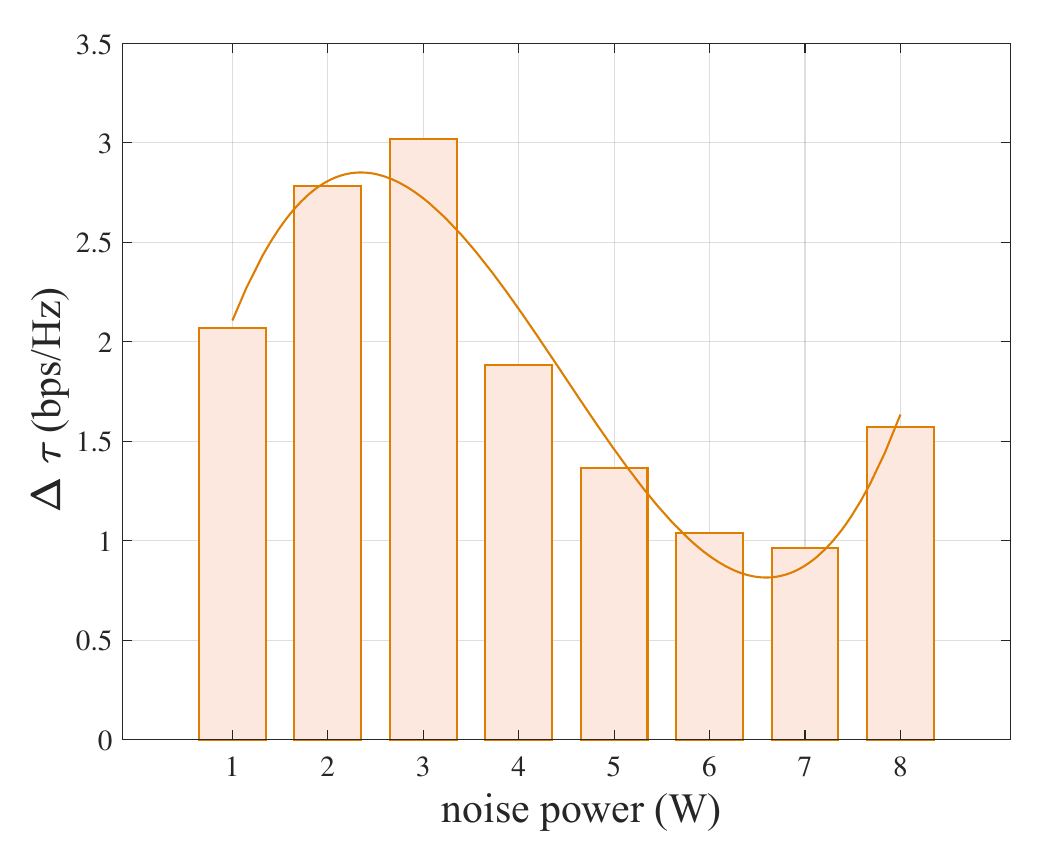}
\caption{\small $\Delta\tau$ \emph{vs.} noise power}
\label{fig:MA-UAV_noise}
\end{figure}

Finally, with the settings of $H=$100m, $M$=4 and $P$=0.1W, Fig. \ref{fig:MA_and_UAV_noise} validates noise power impact on ASR, demonstrating decreasing ASR for both MA micro-mobility and UAV macro-mobility with increasing noise due to Bob's greater sensitivity to noise variations compared to Eve. The MA-based micro-mobility scheme consistently outperforms the UAV-based counterpart in these conditions, as described by fitted curves converging, though Fig. \ref{fig:MA-UAV_noise} reveals the performance gap ($\Delta\tau$) progressively narrows at higher noise levels,  MA's fine-grained spatial control advantage diminishes under severe noise interference as environmental disturbances overwhelm signal refinement capabilities.

\emph{Findings}: All above results demonstrate the complementary advantages of both mobility paradigms: The MA-based micro-mobility approach achieves superior secrecy performance under low transmit power and moderate antenna configurations by leveraging fine-grained spatial reconfigurability to enhance legitimate link quality and suppress signal leakage. Conversely, the UAV-based macro-mobility approach excels in high-power regimes with larger antenna arrays through global mobility that enables optimal positioning relative to legitimate users and eavesdroppers. Strategic integration of MA's local adaptability and UAV's global flexibility can significantly enhance overall secrecy performance.

\begin{figure*}[t]
	\centering
	
	\begin{minipage}[b]{0.48\textwidth}
		\centering
		\begin{subfigure}[b]{0.48\textwidth}
			\includegraphics[width=0.95\textwidth]{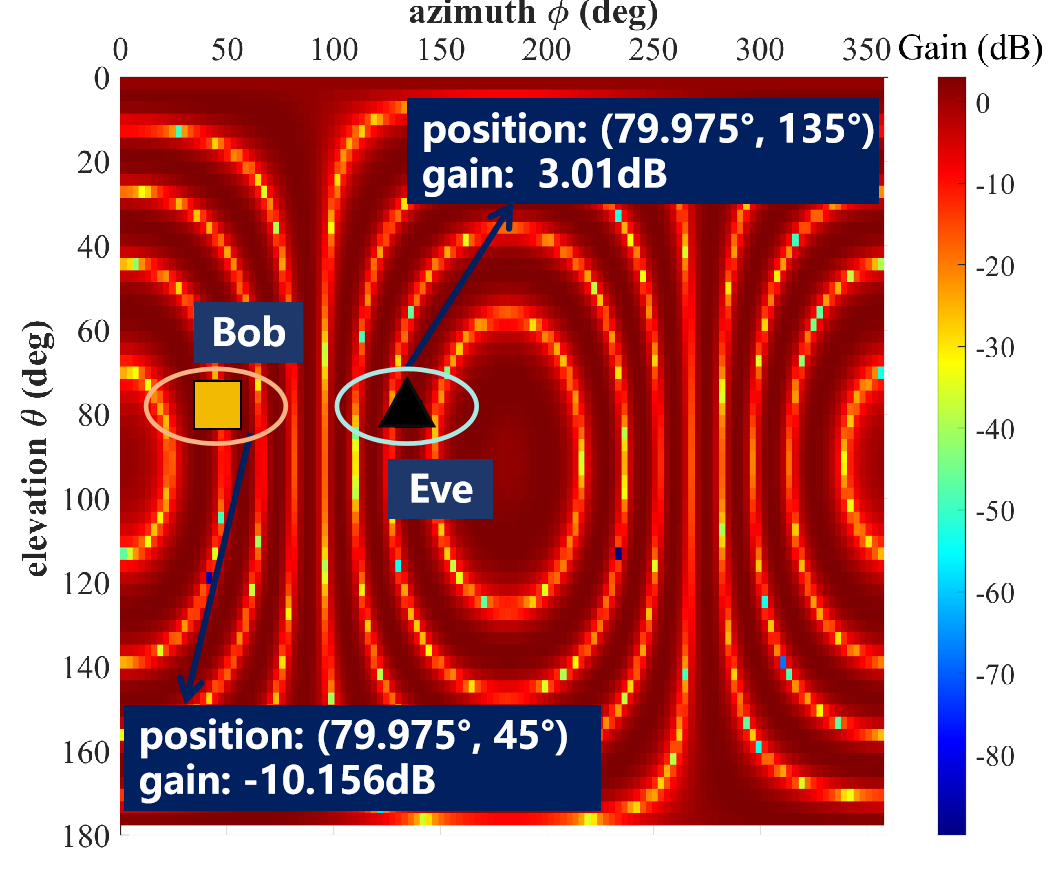}
			\subcaption{Previous beam gain}
			\label{fig:MA_gain1_(M=2)}
		\end{subfigure}
		\hfill
		\begin{subfigure}[b]{0.48\textwidth}
			\includegraphics[width=\textwidth]{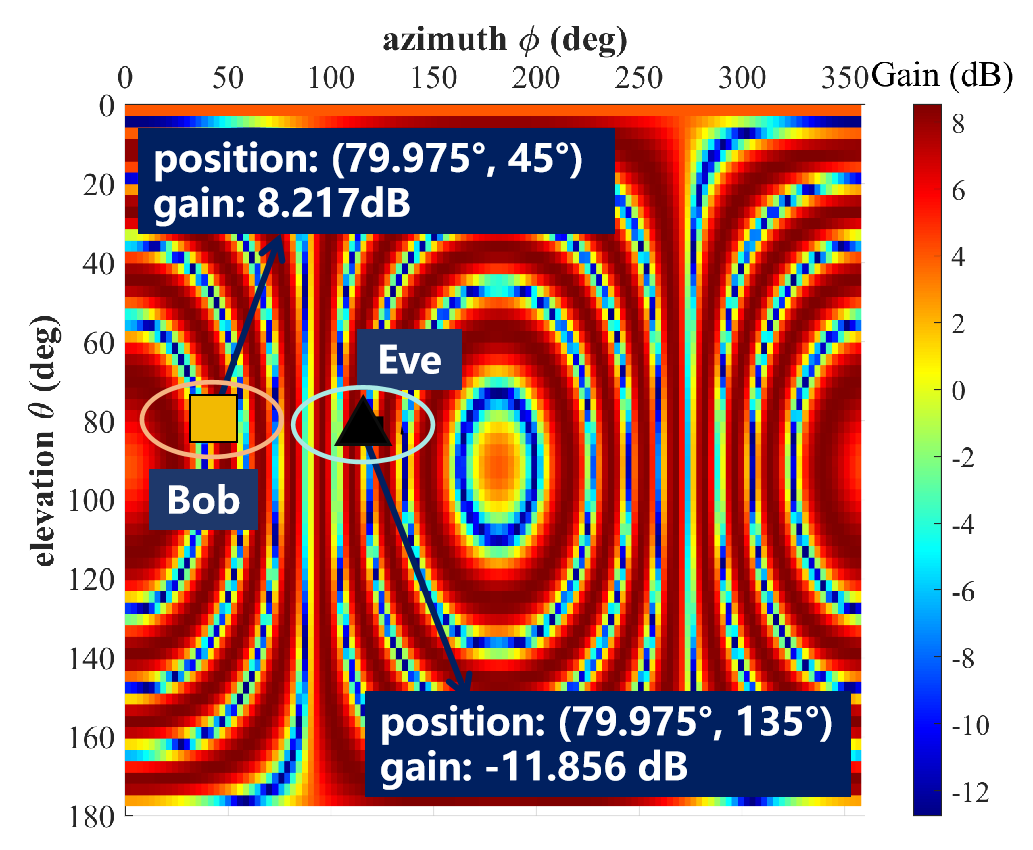}
			\subcaption{Optimized beam gain}
			\label{fig:MA_gain_opt_(M=2)}
		\end{subfigure}
		\caption{Receive antenna gain at different locations ($4\lambda$ space)}
		\label{fig:MA_gain_4}
	\end{minipage}
	\hfill
	\begin{minipage}[b]{0.48\textwidth}
		\centering
		\begin{subfigure}[b]{0.48\textwidth}
			\includegraphics[width=1.1\textwidth]{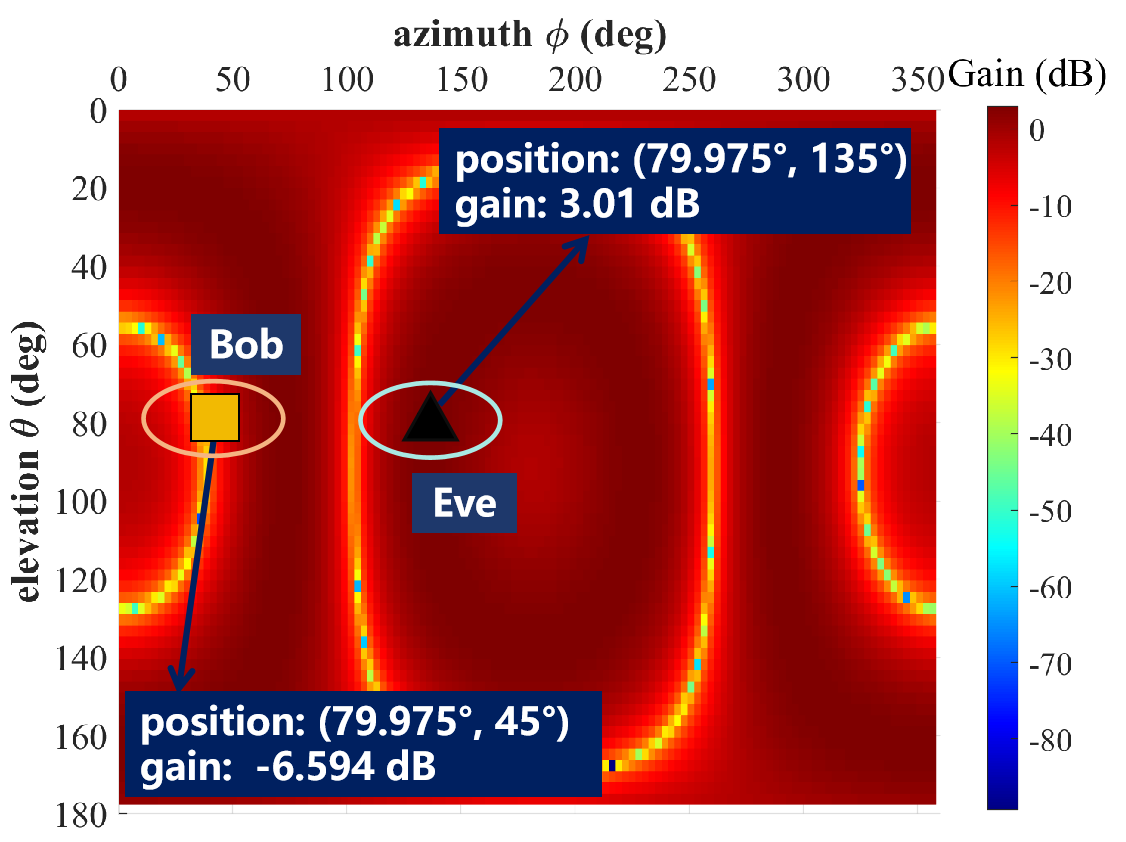}
			\subcaption{Previous beam gain}
			\label{fig:MA_gain1_(M=2)_1lambda}
		\end{subfigure}
		\hfill
		\begin{subfigure}[b]{0.48\textwidth}
			\includegraphics[width=1.02\textwidth]{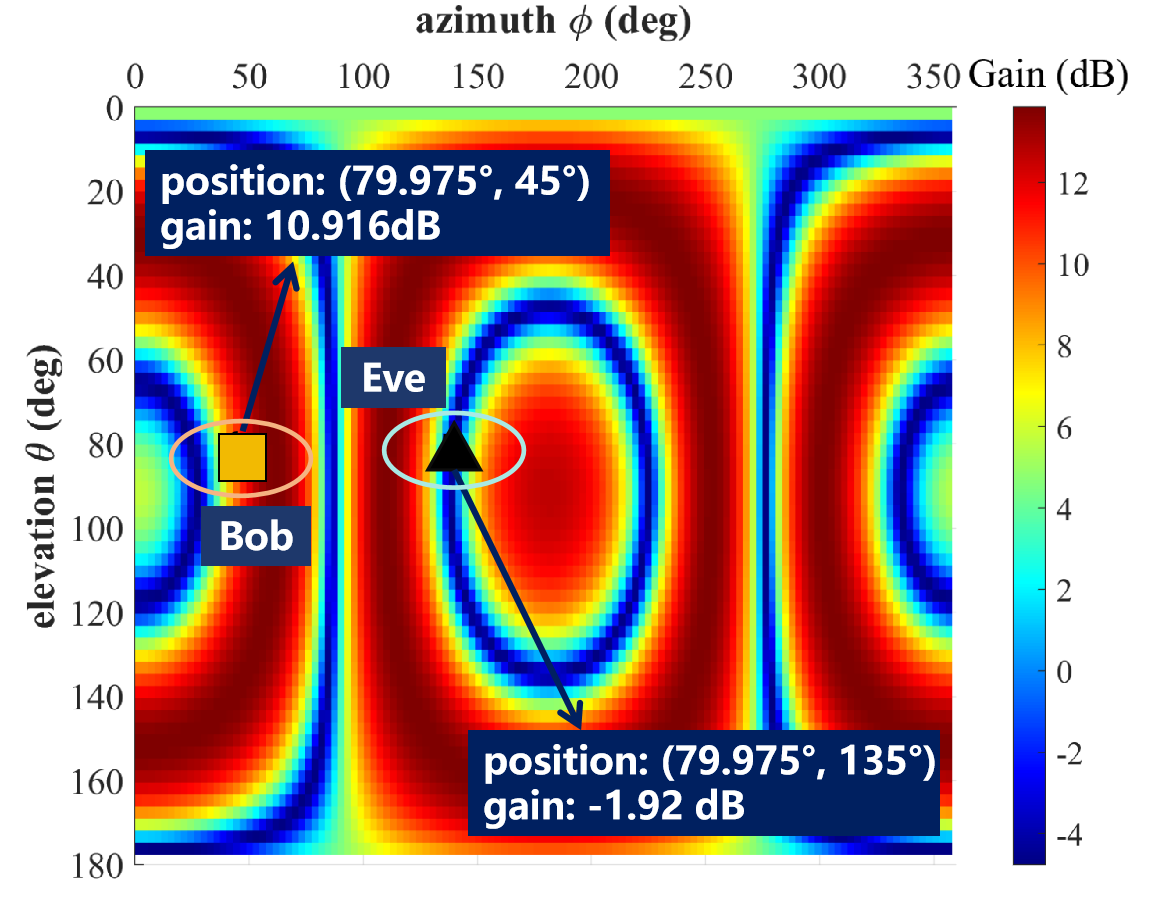}
			\subcaption{Optimized beam gain}
			\label{fig:MA_gain_opt_(M=2)_1lambda}
		\end{subfigure}
		\caption{Receive antenna gain at different locations ($\lambda$ space)}
		\label{fig:MA_gain_1}
	\end{minipage}
	
\end{figure*}

\begin{figure*}[t]
	\centering
	
	\begin{minipage}[b]{0.48\textwidth}
		\centering
		\begin{subfigure}[b]{0.48\textwidth}
			\includegraphics[width=\textwidth]{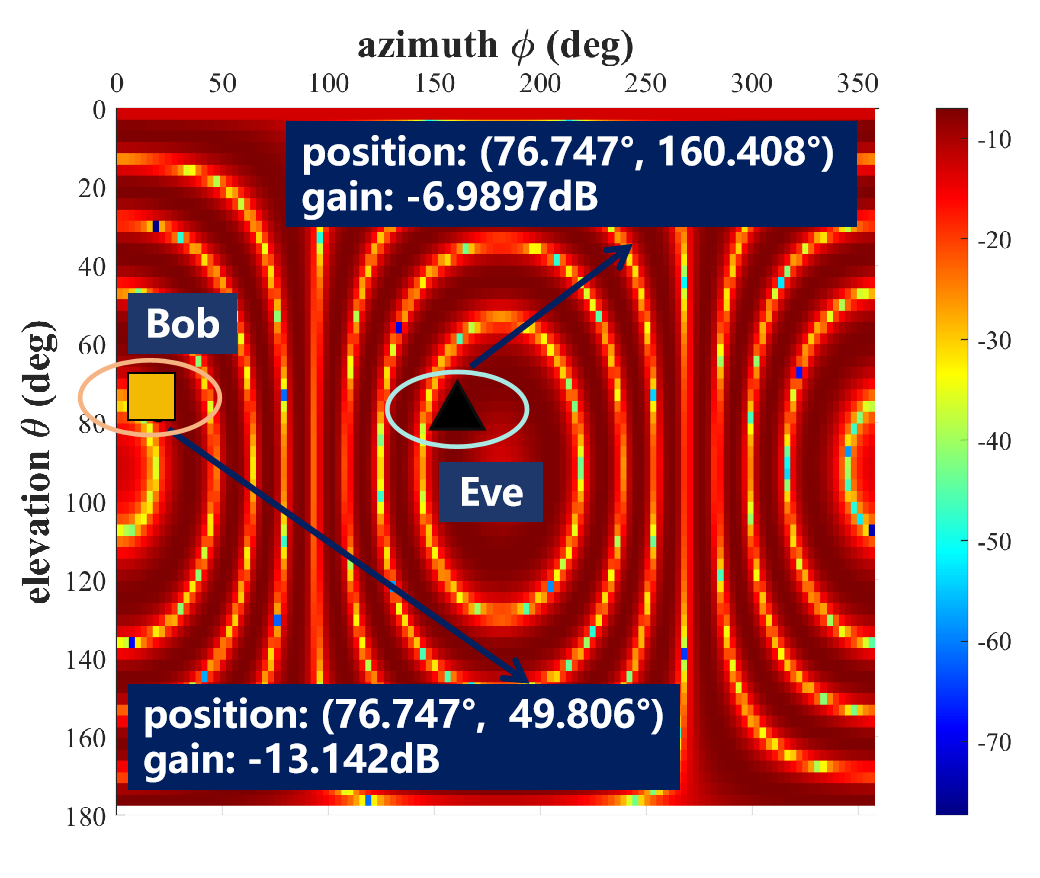}
			\subcaption{Previous beam gain}
			\label{fig:UAV_Gain1(M=2,T=20-(60))}
		\end{subfigure}
		\hfill
		\begin{subfigure}[b]{0.48\textwidth}
			\includegraphics[width=1.03\textwidth]{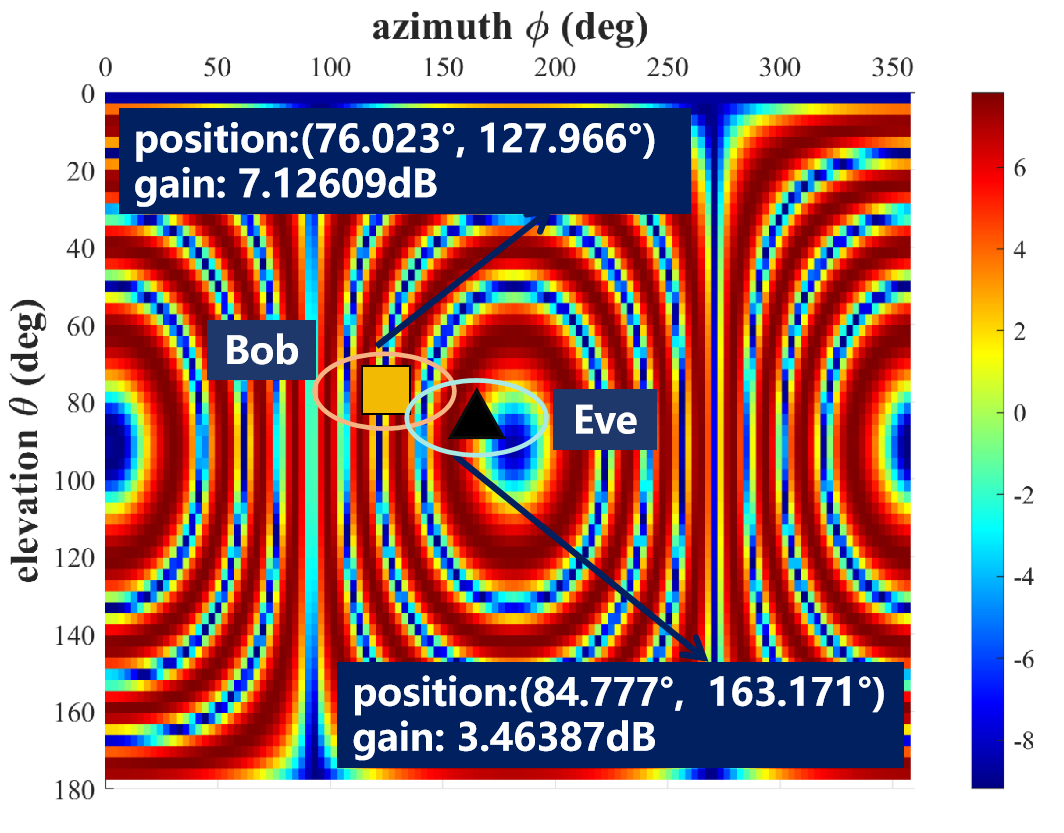}
			\subcaption{Optimized beam gain}
			\label{fig:UAV_Gain_opt(M=2,T=20-(60))}
		\end{subfigure}
		\caption{Receive antenna gain (20-th time slot)}
		\label{fig:UAV_gain_20}
	\end{minipage}
	\hfill
	\begin{minipage}[b]{0.48\textwidth}
		\centering
		\begin{subfigure}[b]{0.48\textwidth}
			\includegraphics[width=1.02\textwidth]{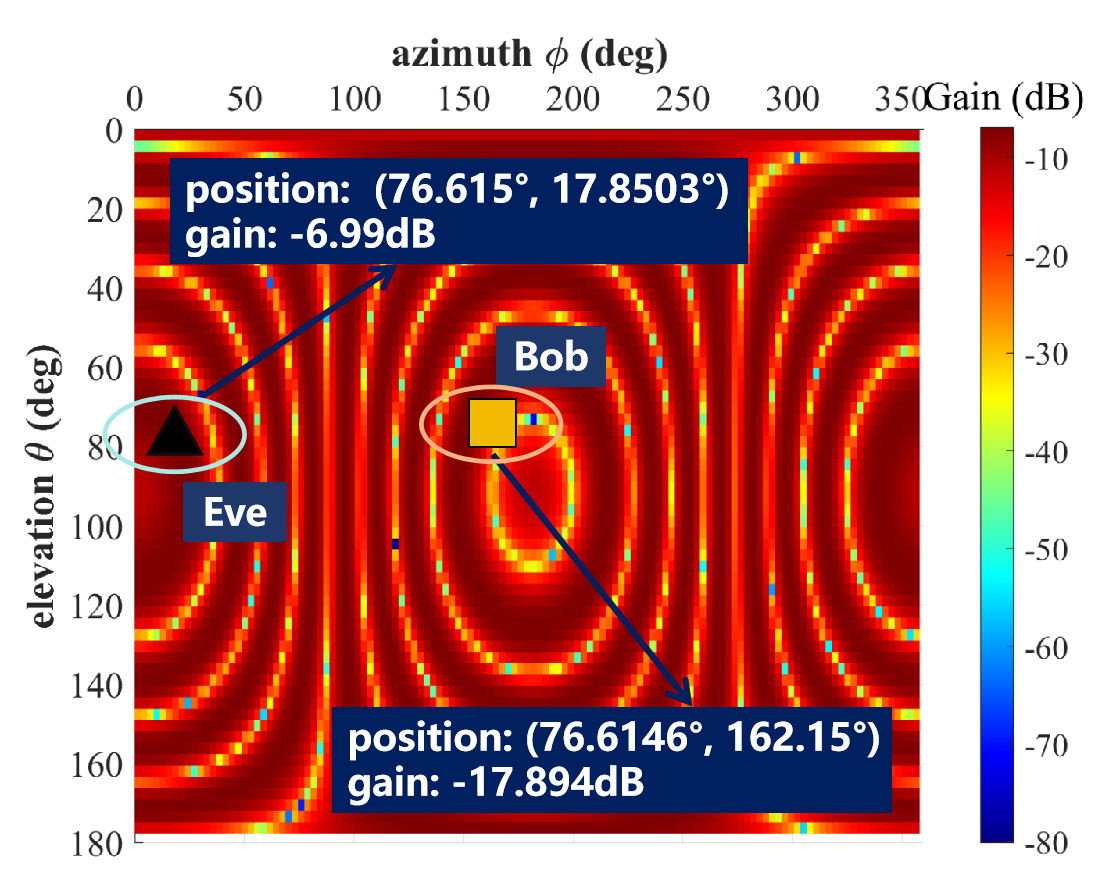}
			\subcaption{Previous beam gain}
			\label{fig:UAV_Gain1(M=2,T=40-(60))}
		\end{subfigure}
		\hfill
		\begin{subfigure}[b]{0.48\textwidth}
			\includegraphics[width=\textwidth]{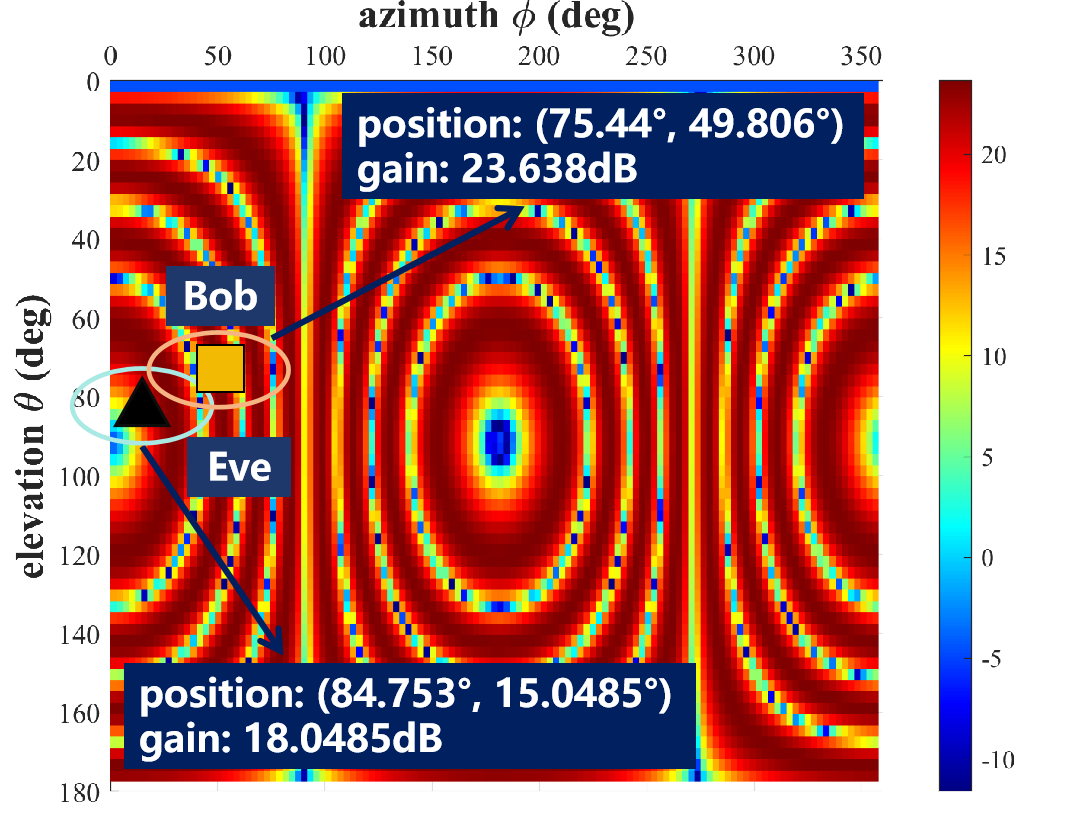}
			\subcaption{Optimized beam gain}
			\label{fig:UAV_Gain_opt(M=2,T=40-(60))}
		\end{subfigure}
		\caption{Receive antenna gain (40-th time slot)}
		\label{fig:UAV_gain_40}
	\end{minipage}
\end{figure*}

\subsection{3D BF gain of MA}
To evaluate BF optimization impacts at Bob and Eve under $P$=1W and $M$=2, each antenna element can move within [0, $4\lambda$], Fig. \ref{fig:MA_gain_4}  demonstrates the MA micro-mobility's 3D gain pattern: no antenna positioning reveals critical security risks with Bob's gain substantially lower than Eve's, as shown in Fig. \ref{fig:MA_gain1_(M=2)}. With the joint optimization of antenna positioning and BF vector, as shown in Fig. \ref{fig:MA_gain_opt_(M=2)}, it achieves 8.217dB gain at Bob and -11.855dB at Eve, yielding a 20.072dB security gap through spatial reconfiguration that concentrates energy toward legitimate users while suppressing Eve leakage.
Contrastingly, when movement range of each MA element decreases from $4\lambda$ to $\lambda$, Fig.~\ref{fig:MA_gain_1} exhibits broader main lobes, smoother angular distributions, and reduced side-lobe fluctuations, indicating degraded spatial resolution from constrained antenna movement. Joint optimization of antenna positioning and BF vector dynamically reshapes radiation patterns to enhance secrecy across varying mobility constraints.

Fig. \ref{fig:UAV_gain_20} and Fig. \ref{fig:UAV_gain_40} demonstrate dynamic BF gain evolution of the UAV macro-mobility in 20-th and 40-th time slots, where non-optimized patterns exhibit poor directional discrimination with energy leakage toward Eve and insufficient gain concentration at Bob across both time instants. While optimized configurations achieve significant spatial reconfiguration: adaptive beam steering precisely tracks Bob's trajectory while suppressing Eve-directed side-lobes by $15$dB at least and enhancing Bob's main-lobe gain by 8-12dB. This temporal adaptability sustains secrecy performance despite nonlinear receiver angle variations during mobility, confirming the framework's capability to dynamically balance signal enhancement and leakage suppression through real-time spatial optimization under UAV macro-scale movement constraints.

\section{Conclusion}\label{sec:conclusion}
Based on the comprehensive investigation conducted in this paper, it was conclusively demonstrated that MA-enabled micro-mobility and UAV-enabled macro-mobility exhibited distinct yet complementary advantages for enhancing PLS in air-to-ground communications. Through the development of a dual-scale mobility framework and the formulation of non-convex ASR maximization problems for both paradigms, the research systematically quantified their performance boundaries. The performance gap analysis further highlighted critical limitations: MA systems faced diminishing returns beyond 5 antennas due to mutual coupling, while UAV systems exhibited restricted flexibility under short mission duration. Environmental factors such as high noise levels and increased flight altitude reduced MA's fine-grained control advantage but minimally affected UAV's macro-scale adaptability. Crucially, neither approach universally superseded the other; instead, their synergistic integration emerged as the most promising direction. The MA's real-time beam manipulation capabilities complemented the UAV's coverage-oriented mobility, suggesting that hybrid micro-macro architectures could optimally balance security, energy efficiency, and deployment complexity in next-generation aerial networks. This work thus established a foundational framework for adaptive mobility strategy selection and pioneered the exploration of hierarchical spatial control in secure wireless communications.

\bibliographystyle{IEEEtran}
\bibliography{ref}
\end{document}